\documentclass[reprint,
superscriptaddress,
nofootinbib, % So you don't have the footnotes at the end but on the same page than initialisation
amsmath,
amssymb, 
aps,
pre,
floatfix,
longbibliography
]{revtex4-2}
\usepackage{makecell}
\usepackage{graphicx}% Include figure files
\usepackage{dcolumn}% Align table columns on decimal point
\usepackage{eqnarray}
\usepackage{subeqnarray}
\usepackage{subcaption}
\usepackage{bm}% bold math
\usepackage{algorithm}
\usepackage{algpseudocode}
\usepackage{hyperref}% add hypertext capabilities
\usepackage{caption}
\usepackage{siunitx}
\usepackage{xcolor}
\usepackage{amsmath,amssymb,amsfonts}
\usepackage{enumitem}
\newcommand{\ind}{\mathbbm{1}}

\usepackage{cancel}

\usepackage{bbm} % dans le préambule

\begin{document}

%\preprint{APS/123-QED}

\title{Wealth Inequality and Planetary Boundaries in a Stylized Agent-Based Model}

\author{Thomas Valade}
\affiliation{Econophysics Lab, Institut Louis Bachelier, 28 Pl. de la Bourse, Palais Brongniart, 75002 Paris, France}
\affiliation{LadHyX UMR CNRS 7646, École Polytechnique, IP Paris, 91128 Palaiseau Cedex, France}
\author{Michael Benzaquen}
\affiliation{Econophysics Lab, Institut Louis Bachelier, 28 Pl. de la Bourse, Palais Brongniart, 75002 Paris, France}
\affiliation{Capital Fund Management, 23 Rue de l’Université, 75007 Paris, France}
\author{Matthieu Cristelli}
\affiliation{Capital Fund Management, 23 Rue de l’Université, 75007 Paris, France}
\author{Stanislao Gualdi}
\affiliation{Capital Fund Management, 23 Rue de l’Université, 75007 Paris, France}
\author{Pierre Lenders}
\affiliation{X-Environnement, Maison des Polytechniciens
12 rue de Poitiers
75007 Paris, France}

\date{\today}

\begin{abstract}
At the intersection of rising wealth inequality and intensifying environmental pressures, we investigate a reverse causal relationship that has received comparatively little attention: wealth inequality may not only be a consequence of environmental crises, but also act as a structural obstacle to the ecological transition itself. We develop a stylized agent-based model in which heterogeneous agents, whose initial wealth follows a Pareto distribution, allocate their income between either a Brown or a Green sector through a utility function. The function is designed to capture the trade-off between short-term returns and exposure to long-term systemic risks. A central ingredient is that wealthier agents perceive themselves as less vulnerable to environmental shocks, thereby reducing the amount of resources available for the transition. We show that, beyond inequality thresholds compatible with those observed in most developed countries, the economy remains locked in a Brown regime, even when a substantial share of agents is sensitive to externalities. We then assess a set of stylized fiscal policies--basic income, carbon taxation, Green incentives, and a combined scheme--and find that their effectiveness depends strongly on the inequality regime and on the regressivity embedded in the fiscal mechanism, revealing multidimensional trade-offs between transition speed, cumulative environmental destruction, growth, and fiscal pressure.
\end{abstract}

\maketitle

%\tableofcontents

\section{Introduction}\label{sec:intro}

Over the past decades, two major structural trends have increasingly attracted the attention of economists and policy makers. The first is the rise in wealth inequality observed within many countries and, in several respects, at the global scale. A growing body of empirical work has documented the increasing concentration of wealth in the hands of the top percentiles of the population (see T. Piketty, G. Zucman et al.~\cite{piketty2006wealth, saez2020rise}). The second trend concerns the intensification of environmental pressures associated with economic activity, including climate change, biodiversity loss, and the depletion of natural resources. These developments have been extensively documented in successive reports of the Intergovernmental Panel on Climate Change~\cite{lee2023ipcc} and the Intergovernmental Science-Policy Platform on Biodiversity and Ecosystem Services~\cite{bba_ipbes_2026}, as well as in scientific frameworks such as the planetary boundaries concept introduced by Johan Rockström and collaborators~\cite{steffen2015planetary, rockstrom2009planetary}.

At the intersection of these two trends, environmental degradation and climate change tend to affect poorer populations disproportionately. Individuals and communities with fewer resources are generally more exposed to environmental shocks and possess fewer means to adapt or recover from them. This relationship has been highlighted in numerous studies, including the influential work of Nicholas Stern on the economics of climate change~\cite{stern2023climate} and more recent analyses of the distributional dimensions of environmental impacts and emissions by scholars such as Lucas Chancel~\cite{chancel2025climate}. More broadly, reflections on the relationship between economic systems and the biosphere—including those of Partha Dasgupta~\cite{dasgupta2021economics}—have emphasized the extent to which contemporary economic growth relies on the degradation of natural capital.
Much of the existing literature therefore examines how environmental crises may exacerbate social inequalities~\cite{diffenbaugh2019global, burke2015global}. 

In this paper, we examine the reverse causal relationship.  
Wealth inequality may not only be a consequence of environmental crises but also act as a structural obstacle to the ecological transition itself.
  More specifically: can inequality generate feedback mechanisms that slow down—or even prevent—the transition toward Greener forms of production and consumption?
Addressing such a question requires tools capable of capturing heterogeneous agents, decentralized decision-making, and the emergence of macro-level dynamics from micro-level interactions.

Traditional economic models based on representative agents and equilibrium assumptions often abstract away from such heterogeneity and feedback loops, making it difficult to study distributional dynamics and systemic transitions simultaneously. For this reason, we adopt an agent-based modeling (ABM) approach. ABMs provide a natural framework for studying complex adaptive systems composed of interacting agents whose aggregate behavior emerges from local rules and interactions. This perspective is closely related to the tradition of complexity economics and econophysics, where “toy” models are widely used to isolate the fundamental mechanisms governing complex phenomena. Foundational works such as Schelling's model of segregation~\cite{schelling1971dynamic} and Epstein \& Axtell’s growing artificial societies~\cite{epstein1996growing}, or Kirman's ants ~\cite{kirman1993ants} showed how simple micro-level rules can produce rich macro-level structures. More recent research uses related heterogeneous-agent approaches to explore monetary policy, instability, and market liquidity (see \cite{gualdi2015tipping}). Their objective is not to reproduce reality in all its details but rather to explore the consequences of a limited set of assumptions and interactions. In the same spirit, the model proposed in this paper should be viewed as a stylized exploratory framework. Its purpose is not to forecast the trajectory of the real-world climate transition, but to investigate potential causal mechanisms and provide a scenario-building tool for examining how different economic and policy configurations may shape systemic outcomes.

Our model seeks to represent an economy composed of a large population of heterogeneous agents whose wealth initially follows a power-law distribution consistent with empirical observations. Economic activity is divided between two sectors: a “Brown” sector associated with carbon-intensive and environmentally harmful activities, and a “Green” sector associated with more sustainable modes of production and consumption. At each time step, agents allocate part of their wealth and income between these two sectors according to a utility function that captures a fundamental trade-off between (i) short-term economic returns, and (ii) exposure to long-term systemic risks generated by environmental degradation. This tension echoes a conflict that has become increasingly visible in public debate—sometimes summarized in France by the expression “fin du mois versus fin du monde,” highlighting the perceived opposition between immediate economic well-being and long-term planetary sustainability.
A key feature of the model concerns heterogeneity in agents’ perceived vulnerability to environmental risks. Wealthier agents may plausibly believe that they are less exposed to such risks, either because they possess diversified assets, enjoy greater geographical mobility, or have the financial means to invest in protective infrastructure. In addition, certain environmental risks can partly be hedged through insurance mechanisms or financial instruments. For the very wealthiest agents, the marginal cost of insuring against additional environmental risks may remain small relative to the additional returns generated by continued participation in the incumbent Brown economy. In such circumstances, it may be individually rational for these agents to prioritize short-term returns and effectively discount long-term environmental risks in their decision-making—not because they are indifferent to environmental outcomes, but because the private cost-benefit calculus they face differs from that of less wealthy agents. This asymmetry between the willingness and the capacity to undertake the transition constitutes one of the central mechanisms we explore here.

While a growing number of studies have used agent-based models to investigate climate and energy transitions~\cite{lamperti2018faraway}, most focus on technological diffusion~\cite{lamperti2018faraway, czupryna2020agent}, policy adoption~\cite{lamperti2019public,safarzynska2022abm,gerdes2022labor}, or consumer behavior~\cite{czupryna2020agent}. In contrast, the model proposed here emphasizes the role of wealth heterogeneity and the interaction between distributional dynamics and environmental risks. In particular, it explores how inequality may influence the probability that an economy transitions endogenously from a predominantly Brown to a predominantly Green configuration. 

The present paper represents a first attempt at modeling how wealth heterogeneity may affect both the capacity and the perceived incentives of agents to support the transition,  and investigate under which conditions a Green transition emerges endogenously. In a second stage of the analysis, we introduce a set of stylized policy interventions—including redistributive mechanisms such as basic income, taxes on Brown economic activities, and incentives for Green investment—in order to evaluate their potential influence on the transition. At this stage, however, the comparative assessment of these policies remains preliminary. The multidimensional nature of the model implies that several distinct criteria may be relevant when evaluating policy outcomes. These include the speed with which a transition toward a Green economy occurs, the cumulative amount of environmental destruction experienced along the transition path, and the scale of redistribution required to facilitate the shift. While these objectives may be correlated, they are not identical, and exploring the trade-offs between them represents an important direction for future research.

The remainder of the paper is organized as follows. Section~\ref{sec:main_model} presents the structure of the agent-based model, including the specification of agents’ utility functions, the representation of the Brown and Green sectors, and the mechanisms through which environmental externalities affect wealth dynamics. Section~\ref{sec:init} describes the baseline behavior of the system and examines the conditions under which a transition toward a predominantly Green economy may occur endogenously. Section~\ref{sec:main_results_nopolicy} introduces a set of stylized policy interventions—including redistributive transfers, taxation of Brown activities, and incentives for Green investments—and evaluates their impact on the likelihood and timing of the transition. Section~\ref{sec:policies} discusses the main results, their limitations, and possible extensions of the model. Section~\ref{sec:conclusion} concludes.

\section{The model}
\label{sec:main_model}

We consider a population of heterogeneous agents $i\in\{1,\dots,N\}$ with non-negative wealth $w_i(t)\ge 0$, where their initial wealth is drawn from a power-law distribution (see Section~\ref{sec:init}). Agents are ranked at $t=0$ from the wealthiest ($i=1$) to the poorest ($i=N$). Time is discrete, with one time step typically corresponding to one year. Agent $i$’s wealth is split between the Brown ($\text{B}$) economy -- high-carbon emitters that damage the planet and contribute to the climate challenge -- and the other part in the Green ($\text{G}$) economy -- low-carbon assets,
\begin{eqnarray}
w_i(t)=w_{i,\mathrm{B}}(t)+w_{i,\mathrm{G}}(t).
\end{eqnarray}
We define aggregate  Brown and Green wealth as
\begin{eqnarray}
W_{\mathrm{B}}(t):=\sum_i w_{i,\mathrm{B}}(t),\qquad 
W_{\mathrm{G}}(t):=\sum_i w_{i,\mathrm{G}}(t),
\end{eqnarray}
and denote $W_{\mathrm{tot}}(t):=W_{\mathrm{B}}(t)+W_{\mathrm{G}}(t)$ the total wealth of agents. In our toy model, both economies bring each agent some immediate satisfaction which could be modeled as "disposable income" (later in the document simply referred to as "income"). The latter is based on their respective Brown and Green wealth, or said otherwise based on their implication into, respectively, business as usual versus highly sustainable modes of consumption and production. Such income varies in time according to the following rates:
\begin{subeqnarray}
\label{eq:returns}
       r_\mathrm{B}(t) &=& r_0 + I\cdot \mathrm{EMA}_{\tau}\left[\frac{W_\mathrm{B}-W_\mathrm{G}}{W_{\mathrm {tot}}}, t\right]  ,\\
   r_\mathrm{G}(t) &=& r_0 - I \cdot \mathrm{EMA}_{\tau}\left[\frac{W_\mathrm{B}-W_\mathrm{G}}{W_{\mathrm {tot}}}, t\right]  ,
\end{subeqnarray}
where $r_0$ is a constant reference return. The exponential moving average $\mathrm{EMA}_{\tau}$ (with time scale $\tau$) sets a proxy for the imbalance between aggregate Brown and Green segments of the economy, which means current investment trends can lead to higher future returns, reinforcing themselves over time.\footnote{The $\mathrm{EMA}$ is defined by $\mathrm{EMA}_{\tau}[f,t] = \sum_{t'=0}^{t} e^{-\frac{t-t'}{\tau}}f(t')$, or through the recursive formula: $\mathrm{EMA}_{\tau}[f,t] = \left(1- \frac{2}{\tau+1} \right)\mathrm{EMA}_{\tau}[f,t-1] + \frac{2}{\tau+1} f(t)$. 
}In the following we shall  use the terms returns, yields, or rates interchangeably for $r_\mathrm{G}$, $r_\mathrm{B}$, and $r_0$.

This term can be interpreted as a stylized representation of mechanisms such as lobbying power, economies of scale or institutional inertia. The parameter $I>0$ controls the strength of this feedback.
The income of an agent is denoted by $y_i(t)$ and reads 
\begin{eqnarray}
y_i(t)= r_{\mathrm B}(t)w_{i,\mathrm B}(t) + r_{\mathrm G}(t)w_{i,\mathrm G}(t).
\end{eqnarray}
The total income of the economy is thus given by $Y_\mathrm{tot}(t)=\sum_iy_i(t)$.

\begin{figure}[t!]
    \includegraphics[width=0.9\columnwidth]{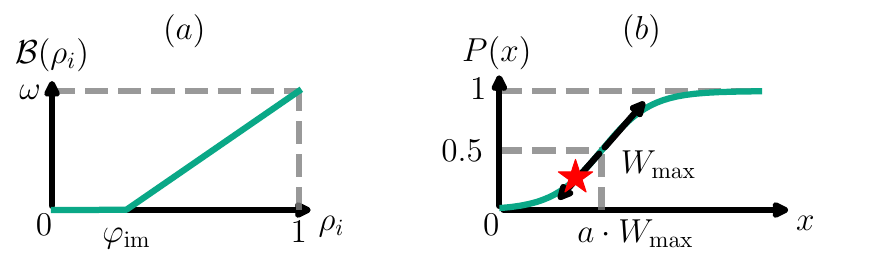}
    \caption{\textit{Behavioral factor} of agents (left) and probability function of a shock induced by climate negative externalities (right). The \textit{red} star corresponds to the baseline situation used in Table.~\ref{tab:parameters}.}
    \label{fig:functions}
\end{figure}

At each time-step, agents decide if they will allocate the income generated at the previous step more towards Green or towards Brown activities. They do so rationally\footnote{The case of fully rational agents can be seen as the limit case of a more general rule of agents with bounded rationality. In this case if we note $\beta$ the rationality term the logit rule would simply be written as follows: $\mathbb{P}(\mathrm G|i,t) = \frac{1}{1+e^{-\beta \Delta u_i(t)}}.$ We leave this interesting extension for future analysis.}, according to the following rule
\begin{eqnarray}
     \mathbb{P}(\mathrm G|i,t) = \ind_{\{ \Delta u_i(t) > 0\}},
\end{eqnarray}
where $\mathbb{P}(\mathrm G|i,t)$ denotes the probability to choose Green over Brown 
and $\Delta u_i$ is the utility difference between the Green and Brown strategies, as computed by agent $i$.\footnote{More precisely $\ind_{\{ \Delta u_i(t) > 0\}}=1$ when $\Delta u_i(t) > 0$, else $0$.} 
Agents measure their utility as a trade-off between instantaneous gains and potential losses induced by their choices, as follows: 
\begin{eqnarray}
 \Delta u_{i}(t) =  (1-\lambda) \Delta M_i(t) - \lambda \mathcal{B}_i(t)  \Delta C_i(t).
 \label{eq:utility_function}
\end{eqnarray}
Equation~\eqref{eq:utility_function} is the backbone of our study. The first term in the RHS measures the difference of income generated by Brown and Green allocations at time $t$, and can be written as
\begin{eqnarray}
\label{eq:LHS}
 \Delta M_i(t)= (r_{\mathrm G}(t)-r_{\mathrm B}(t)) \frac{y_i(t)}{Y_{\mathrm {tot}}(t)}.
 \end{eqnarray}
The second term in the RHS of Eq.~\eqref{eq:utility_function} instead accounts for the difference in average future wealth destruction induced by preferring Brown over Green (thereby increasing negative externalities), and is the product of three terms.
The first, the parameter $\lambda$, sets the global \textit{negative externalities awareness}: i.e. the overall relative weight of the two contributions. The second aims at encoding the heterogeneity of agents' sensitivities via a behavioral factor $\mathcal{B}_i$. 
We assume that $\mathcal{B}_i=\mathcal{B} (\rho_i)$ depends only on the rank $\rho_i(t)$ of the agent in the wealth distribution, as
\begin{eqnarray}
\label{eq:behavior}
    \mathcal{B} (\rho_i) = \left\{ \begin{array}{cl}
0 & , \ \rho_i \leq \varphi_\mathrm{im} \\
\omega \frac{\rho_i-\varphi_\mathrm{im}}{1 - \varphi_\mathrm{im}} & , \ \rho_i  > \varphi_\mathrm{im}
\end{array} \right.
\end{eqnarray}
where $\omega$ is a normalization constant,\footnote{More precisely, such normalization allows to have a coherent scale between both components of the utility function. To compute $\omega$, we take the reference initial conditions and parameters shown in Table~\ref{tab:parameters}, and  compute $\Delta M_{N/2}$ and $\frac{1}{2}\Delta C_{N/2}$ for the median agent at $t=0$, and take $\omega = 2 \times {\Delta M_{N/2}}/{\Delta C_{N/2}}$.} $\varphi_\mathrm{im} \in [0,1)$ 
denotes the fraction of the richest agents who feel immune to negative externalities (see Fig.~\ref{fig:functions}a for a graphical representation).\footnote{The shape of this function emerges from the observation that the wealthier an agent is, the more access they have to hedging strategies and the more capable they are of adapting to negative externalities.} 
Equation~\eqref{eq:behavior} encodes in a stylized fashion the existing link between willingness to pay for environmental protection and income, with a decreasing trend among high-income earners, see e.g.~\cite{shao2018rich}. %Here, we have opted for a simplified version featuring a purely decreasing relationship, as lower-income earners have only a marginal influence on the transition.

The third term $\Delta C_i$ is the \textit{expected cost difference} induced by the non-sustainable component of investing in the Brown economy. More specifically, the rate of environmental shocks depends on the amount of wealth invested in Brown assets and we model the probability for a shock to happen at a time $t$ as:
\begin{eqnarray}
\label{eq:pshock}
    P\left(x\right) = \frac{1}{2} \left[ 1 + \mathrm{tanh}\left(\frac {x} {W_\mathrm{max}} -a\right) \right]
\end{eqnarray}
where $W_\mathrm{max}$ is a proxy of the maximum amount of Brown wealth the Earth can handle and $a$ is a constant that defines the point at which the rate of shocks accelerates significantly (see Fig.~\ref{fig:functions}b for a graphical representation). The latter captures the concept of a climate tipping point, that has been used widely in the literature (see ~\cite{antilla2010self,lenton2008tipping}). Replacing $x$ by 
$\mathrm{EMA}_{\theta}[W_\mathrm{B}]$ gives the probability that negative externalities materialize at each time step. The $\mathrm{EMA}_{\theta}$ encodes the fact that negative externalities have an impact over a typical timescale $\theta$.

The expected cost difference mentioned above thus reads $\Delta C_i(t) =$
{\small
\begin{eqnarray}
\label{eq:RHS}
     - r_{\mathrm{loss}} \Bigg[&& P\bigg(\Big(1-\frac{2}{\theta+1}\Big)\mathrm{EMA}_{\theta}[W_{\mathrm B},t] +\frac{2}{\theta+1}(y_i(t)+W_\mathrm{B}(t))\bigg) \nonumber\\
    &&- P\bigg(\Big(1-\frac{2}{\theta+1}\Big)\mathrm{EMA}_{\theta}[W_{\mathrm B},t] +\frac{2}{\theta+1}W_\mathrm{B}(t)\bigg) \Bigg]
\end{eqnarray}}
where $r_{\mathrm{loss}}$ is the average  rate of loss in wealth due to externalities related destruction.\footnote{This formulation encodes the change in the probability of a shock from an agent’s perspective, depending on whether the agent allocates previously earned income towards Brown or Green.}
The effects of such negative externalities are modeled by shocks induced by a random multiplicative loss with rate $r_{\mathrm{loss}}$.\footnote{For the sake of simplicity, we chose not to make a dependence between $w_i$ and  $r_{\mathrm{loss}}$, even though such hypothesis has been recently challenged by Hallegatte et al.~\cite{hallegatte2018poverty}.} 
A shock materializes as a global loss of wealth:  with probability $P(\mathrm{EMA}_{\theta}[W_{\mathrm B},t])$ (see Eq.~\eqref{eq:RHS}) each agent loses a random fraction $L_i(t)$ of its wealth, drawn from a uniform distribution $\in[0, 2r_\mathrm{loss}]$, while with probability $1-P(\mathrm{EMA}_{\theta}[W_{\mathrm B},t])$ the agent's wealth is left unaffected. We also introduce a yearly amortization or depreciation component $a_\mathrm{G}$ and $a_\mathrm{B}$, acting resp. on both Green and Brown wealth.\footnote{This is equivalent to reducing the average yield $r_0$.}\\

\begin{figure}[t!]
    \includegraphics[width=0.5\textwidth]{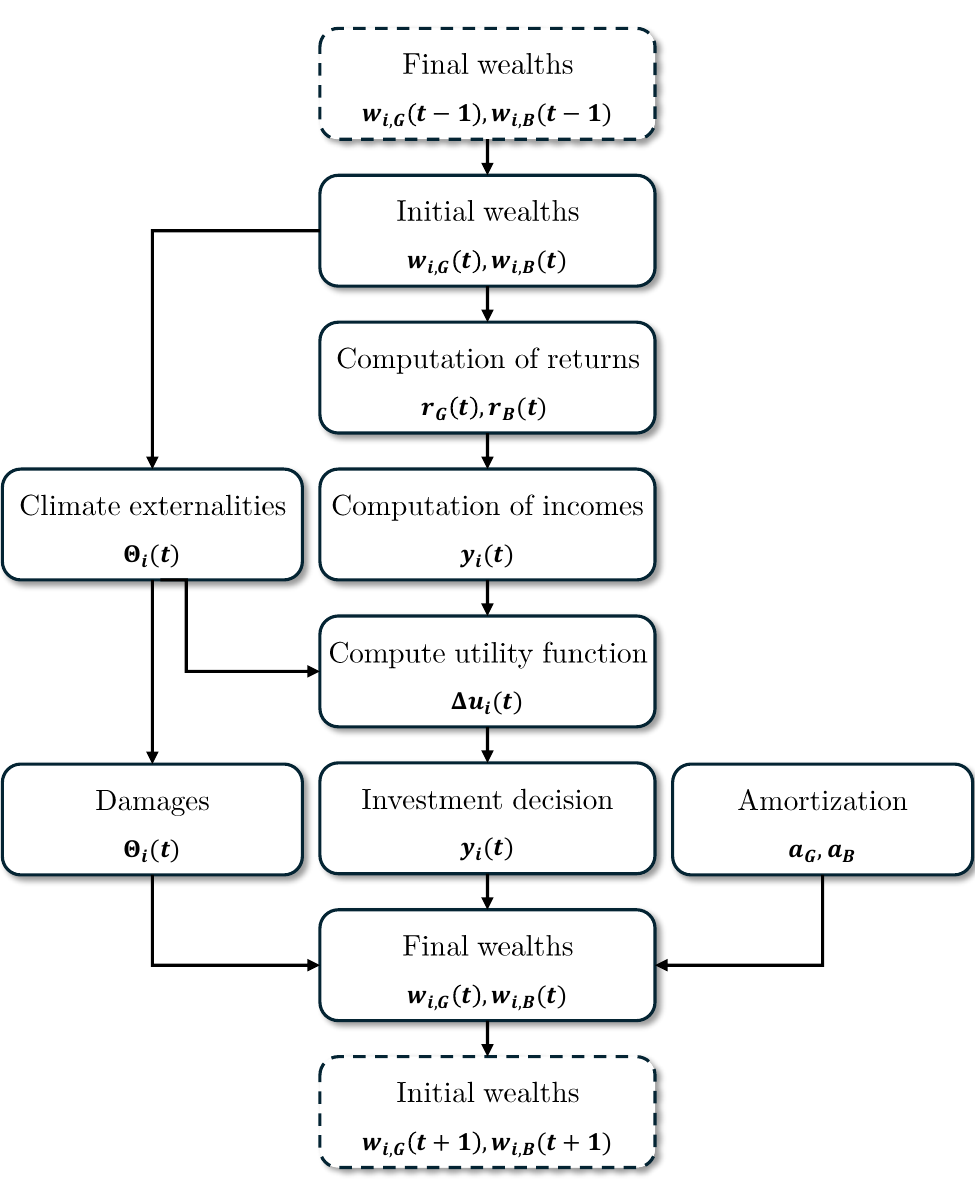}
    \caption{Global workflow of the model.}
    \label{fig:model}
\end{figure}

In the end, the wealth of each agent evolves according to the following rules. If  agent $i$ chooses to invest in Green activities:
\begin{subequations}
\small
    \begin{eqnarray}
     \left\{ \begin{matrix}
w_{i,\mathrm{G}}(t+1) =& \left[ w_{i,\mathrm{G}}(t)\cdot(1-a_\mathrm{G}-\Theta_i(t))+ y_i(t)\right] \\
w_{i,\mathrm{B}}(t+1) =&  \left[w_{i,\mathrm{B}}(t)\cdot(1-a_\mathrm{B}-\Theta_i(t))\right]
\end{matrix} \right. 
\end{eqnarray}
If, on the other hand, an agent chooses to invest in Brown activities:
\begin{eqnarray}
    \left\{ \begin{matrix}
    w_{i,\mathrm{G}}(t+1)=& \left[w_{i,\mathrm{G}}(t)\cdot(1-a_\mathrm{G}-\Theta_i(t))\right]\\
w_{i,\mathrm{B}}(t+1)=& \left[w_{i,\mathrm{B}}(t)\cdot(1-a_\mathrm{B}-\Theta_i(t))+y_i(t) \right]
\end{matrix} \right.
\end{eqnarray}
\end{subequations}
A summary scheme of the model is depicted in Fig.~\ref{fig:model} (see also Appendix~\ref{Pseudo_code} for the detailed pseudo-code).

\section{Model initialization and parametrization}\label{sec:init}
\subsection{Initial condition for wealth heterogeneity}
The degree of heterogeneity in agents' wealth will play a critical role in shaping the dynamics of the model \textit{vis-à-vis} the inhibition/enhancing of the transition from Brown-to-Green economy. We will adopt a phenomenological model of wealth distribution. This will allow us i) on the one hand to make the initial condition adhere to the main empirical features available in the literature and ii) on the other hand to derive an actionable relationship between the parameters of our modeling and a well known concentration indicator, i.e. the Gini coefficient. 
Namely, the upper tail of the wealth distribution is empirically well fitted by a Type-II Pareto distribution~\cite{gabaix2009power} with parameter $\mu=0$:
\begin{eqnarray}
    \mathcal{S}(w, k, \ell) = \left(\frac{\ell}{w+\ell}\right)^k \ , 
\end{eqnarray}
where $\mathcal{S}(w, k, \ell)$ denotes the survival function
with $k$ the exponent of the Pareto law and $\ell$ the scale parameter.
 The corresponding initial wealth of the agent of rank $\rho_i = \frac{i}{N}$ with $i \in [\![1,N ]\!]$ (from the  wealthiest to the poorest) is given by:
\begin{align}
\label{eq:initial_wealth}
    w_i(t=0) &=\ell \int^{1-\rho_{i-1}}_{1-\rho_i}\mathcal{W}(p')dp' \nonumber \\
    &= \ell\frac{k}{1-k}\left[\rho_{i-1}^{1-\frac{1}{k}}-\rho_i^{1-\frac{1}{k}}\right]-\frac{\ell}{N}
\end{align}
where $\mathcal{W}(p)$ is the generalized inverse of the survival function $\mathcal{S}$ taken in $1-p$:IO 
\begin{eqnarray}
\label{eq:initial_wealth_th}
    \mathcal{W}(p)=\ell\left(\frac{1}{(1-p)^\frac{1}{k}}-1\right).
\end{eqnarray}
 This formulation leads to a theoretical Gini coefficient  \begin{eqnarray}Gini(k)=\frac{k}{2k-1}.\end{eqnarray}
 We also define the share $w_\Delta$ of wealth held by the  fraction $\Delta$ of richest agents:
\begin{eqnarray}w_\Delta=\frac{W_\Delta}{W_{tot}}=\Delta \left(\frac{k}{\Delta^\frac{1}{k}}-(k -1) \right),\end{eqnarray}
with $W_\Delta$ the wealth held by the fraction $\Delta$ of wealthiest agents. 
Following the work of~\citet{vermeulen2018fat}, we aggregate wealth data using Consumer \& Finance surveys, and Forbes' billionaires list. The analysis gives estimates of $k$ and $\ell \sim10^5$  (see Appendix~\ref{sec:initial_wealths} and Fig.~\ref{fig:wealth_empirical}). 

\subsection{Model parameters}
We assume that at $t=0$, all agents have the same share $ratio_\mathrm{G}$ of Green to Brown wealth.
The model exhibits similar results in a wide range of values for the number of agents (from  $N=1000$ to the maximum value we tested $N=10^7$). The chosen value $N=1000$ is a trade-off between the computational time and a value for $N$ large enough to avoid the finite size effects due to this parameter.
For several of the model parameters, although not explicitly calibrated, we have followed a \textit{plausibility/realism} approach: i.e. we have chosen values which can be related to empirical values or/and are in the range of empirical values under the hypothesis -- as previously discussed -- that one time step corresponds to one calendar year. 
Amortization $a_\mathrm{B}$ and $a_\mathrm{G}$ are set to reflect the depreciation of assets over a time between 10 and 20 years. Half of the maximum spread $I$ of the economy yields and $r_0$ are chosen to represent a realistic order of values for yields deriving from two types of economic investments. The $\theta$ represents the time scale for the decay of the impact of the externalities and we set it to be close to the value which is typically indicated for greenhouse gases. The other time scale of the model $\tau$ can be intended as a friction due to interplay of several factors such as lobbying, scale effects, time to deploy technological developments, etc. It is set to be in the same range of the time of amortization. The amortization time indeed can be seen as the typical time at which physical assets become obsolete and $\tau$ as the time required for changes to get full scale. We refer to Table~\ref{tab:parameters} for the value of the remaining reference parameters. In what follows, we will calculate statistics on the simulations based either on the means or on the medians, whichever is more appropriate. The mean of the quantity $A$ at time $t$ over the runs will be denoted by $\langle A(t)\rangle$, and the median will be denoted by $\tilde{A}$.

\begin{table}[t!]
\caption{\label{tab:parameters}Reference values of parameters.}
\begin{ruledtabular}
\begin{tabular}{lll}
\textbf{Parameter} & \textbf{Notation} & \textbf{Value}\\
\hline 
Number of agents & $N$ & 1000-10000\\
Negative externalities awareness & $\lambda$ & $50\%$ \\
Max $W_\mathrm{B}$ Earth can support & $W_\mathrm{max}$ & $100$\\
 \makecell[l]{EMA decay factor \\ \quad of negative externalities} & $\theta$ & $100 \text{ years}$\\
Tech inertia & $\tau$ & $5 \text{ years}$\\
Average yield extracted from wealth& $r_0$ & $7\%$\\
\makecell[l]{Max difference \\ \quad between $r_\mathrm{G}$ or $r_\mathrm{B}$, and $r_0$} & $I$ & $5\%$ \\
Expected damage in case of loss& $r_\mathrm{loss}$ & $10\%$\\
Inflection point & $a$ & $2.15$ \\
Natural amortization \& depreciation & $a_\mathrm{B} \ or \ a_\mathrm{G}$ & $5\%$ \\
\makecell[l]{Fraction of agents \\ \quad rich enough to feel immunized} & $\varphi_\mathrm{im}$ & $0.1\%$  \\
Normalization constant & $\omega$ & $\sim 2 \cdot 10^4$ \\
Maximum number of timesteps & $t_\mathrm{max}$ & 100 years \\

\hline 
\makecell[l]{\textit{Policy-specific parameters} \\ \quad \textit{(used in the last section)}}\\
\hline 
Max. level of taxes on incomes & $r_\mathrm{tax}$ & $10\%$ \\
Tax-rate cap multiplier & $\alpha_\mathrm{min}$ & $10\%$ \\
\makecell[l]{Regressivity threshold \\ \quad(income/median)} & $q_1$ & $20$ \\
Tax regressivity slope parameter & $q_2$ & $100$
\end{tabular}
\end{ruledtabular}

\caption*{\label{tab:variables}Reference values of initial conditions.}
\begin{ruledtabular}
\begin{tabular}{lll}
\textbf{Initial condition} & \textbf{Notation} & \textbf{Value}\\
\hline 
Initial total wealth & $W_\mathrm{tot}(0)$ & $1.7 \cdot W_\mathrm{max}$\\
Initial ratio of $W_\mathrm{G}$ & $ratio_\mathrm{G}$ &  $15\%$\\
Initial Gini of wealths & $Gini_{0}$ & 0.75-0.85

\end{tabular}
\end{ruledtabular}

\end{table}
\section{RESULTS}
\label{sec:main_results_nopolicy}
\subsection{Baseline configuration}

We first study the model in a baseline configuration without climate-mitigation policy to characterize its dynamics and discuss its main characteristics. We define the end-state of the system as the state reached after at $t=t_\mathrm{max}$ and the outcome for this state is binary: (i) \emph{transitioned} if $r_\mathrm{G}(t_\mathrm{max}) >r_\mathrm{B}(t_\mathrm{max})$,  (ii) \emph{not-transitioned} otherwise. By construction of the model, once the system has transitioned to a G-state (i.e., $r_\mathrm{G}(t)>r_\mathrm{B}(t)$), it cannot \textit{de facto} revert to a B-state (i.e., $r_\mathrm{B}(t)>r_\mathrm{G}(t)$) because the agents do not have any incentive to reinvest in Brown assets (see Eq.~\eqref{eq:LHS}).\footnote{Up to fluctuations due to climate shocks.}We define the time to transition $t=T_2T$ as the first time we observe the inversion of the two yields, e.g. $r_\mathrm{G}>r_\mathrm{B}$.

\subsection{Phase diagrams}
\label{sec:phasediagram_nopolicy}
The first dimension we want to explore as a function of the parameter and initial condition space is whether the system can undergo the transition from a Brown to a Green economy and in how much time. In particular we test the dependence of the $T_2T$ and the final outcome of the model as a function of six key dimensions/parameters: 
\begin{enumerate}[label=\roman*]
    \item the initial level of wealth heterogeneity $Gini_{0}$,
    \item the initial share of Green-wealth owned by each agent $ratio_\mathrm{G}$ (i.e. how far the system is from the necessary condition unlocking a higher yield of Green economy against Brown economy), 
    \item the average damage when a shock occurs $r_\mathrm{loss}$,
    \item the negative externalities awareness $\lambda$,
    \item the fraction of the richest agents who consider themselves insulated from the effects of climate shocks $\varphi_\mathrm{im}$ (those agents are consequently driven only by the first term of the utility function in Eq.~\eqref{eq:utility_function}),
    \item and the maximum spread $I$ of the returns of Green and Brown economy from $r_0$.
\end{enumerate}

\begin{figure*}[t]

  \begin{subfigure}{0.49\textwidth}
  
    \includegraphics[width=\textwidth]{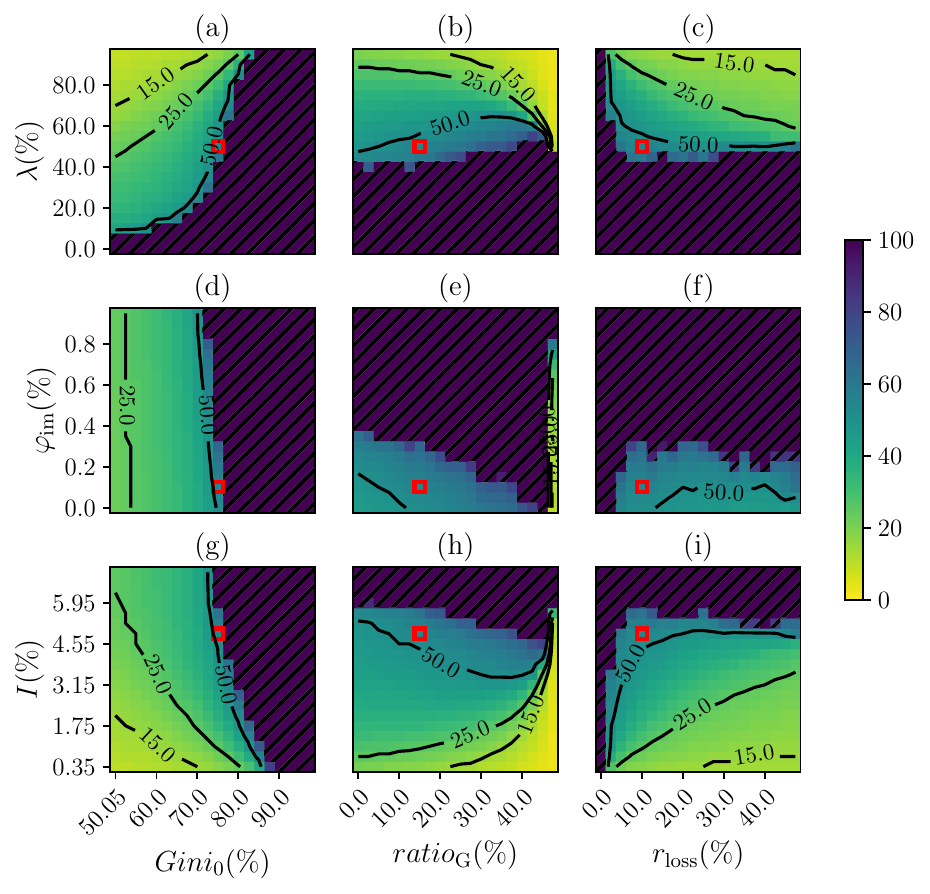}
    \caption{Initial $Gini_{0}=0.75$}
    \label{fig:giniratio1}
  \end{subfigure}
  \hfill
  \begin{subfigure}{0.49\textwidth}
    \includegraphics[width=\textwidth]{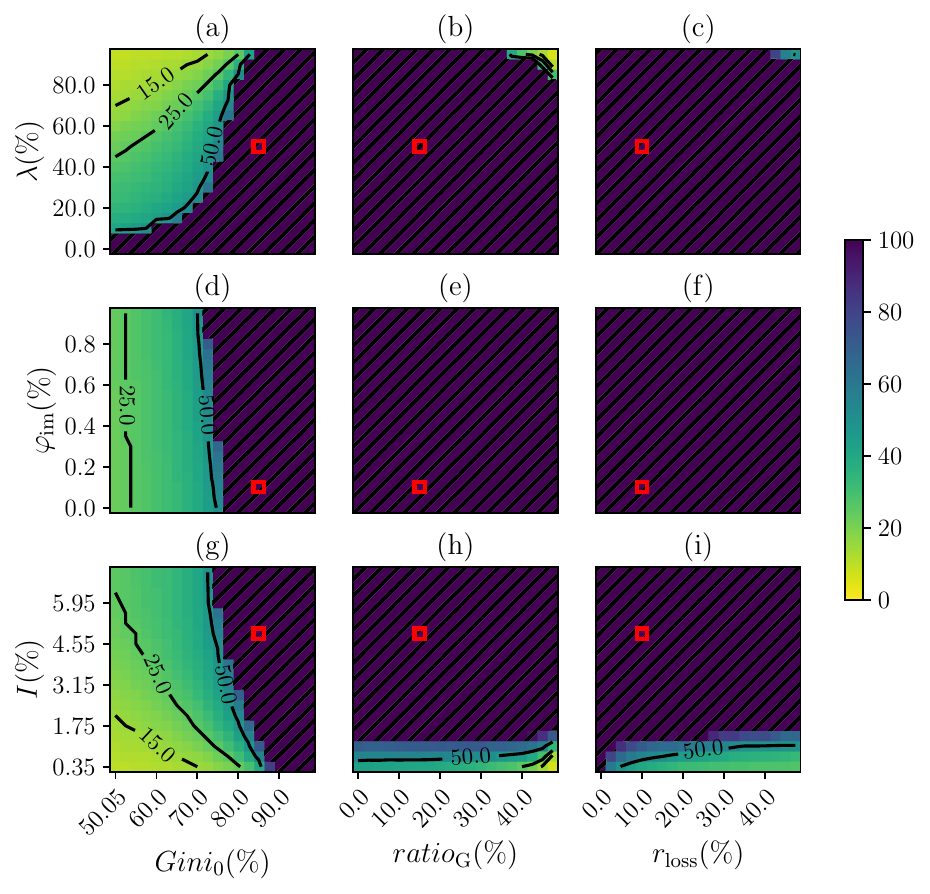}
    \caption{Initial $Gini_{0}=0.85$}
    \label{fig:giniratio2}
  \end{subfigure}
  
  \caption{Phase diagrams of the median transition time ($T_2T$) from Monte Carlo experiments (100 simulations). The hatched region indicates parameter combinations for which at least half of the simulations fail to transition. Contour lines show iso-$T_2T$ levels (equal median $T_2T$). The red square indicates a point with the same parameters (i.e., reference parameters) in all the figures. It should be noted that the first column remains the same in both figures because the $Gini_0$ coefficient is used as the parameter of analysis in these columns.}
  \label{fig:giniratio}
\end{figure*}

In Fig.~\ref{fig:giniratio} we show how these parameters affect the time to transition ($T_2T$) for two reference values of the $Gini_0$ for the initial wealth, namely $0.75$ and $0.85$. Although agent-based models are typically not calibrated (or cannot be easily calibrated) on empirical time series/regimes, we have designed the model to have \textit{plausible} parameter ranges. This allows a \textit{prudent} link between the model initialization, scenarios and empirical observations. It will guide us in the exploration of the parameter space and in the interpretation of the results. As an example, the range encompassed by the two chosen values for $Gini_0$ appears consistent with empirical observations for most developed countries (see Appendix.~\ref{sec:initial_wealths}). We stress again that we are not claiming we can calibrate the model on the specific economic state of a country. We work under a weaker hypothesis that several of the model parameters have plausible value ranges and can produce scenarios providing insights on real economic systems and potentially supporting specific policy recommendation.\\ 

Across multiple simulations and setups of the model, transitions occurring after 100 years were found to be unlikely. Similarly, across multiple runs for a fixed parameter set, transitions beyond $t_{\max}$ were also unlikely, which justifies setting this parameter to $t_{\max}=100$. In Fig.~\ref{fig:giniratio1} and Fig.~\ref{fig:giniratio2}, the darker the heatmap, the longer the $T_2T$ measured and no transition is observed in the hatched regions. Similarly to $Gini_0$, we remind that the unit time step of the model is not explicitly calibrated but it is representative of one calendar year. This implies that for the tractability of our results, we consider that under a sufficient number of runs there is no transition after $100$ years. For heterogeneity levels compatible with most developed countries, the model suggests that in the best case scenario (Fig.~\ref{fig:giniratio1}) we are in a regime where the transition is possible for some regions of the parameter space and in the worst case scenario (Fig.~\ref{fig:giniratio2})) we are in a system locked in the Brown state regardless of the values of the other parameters. Let us now focus on the panels (e) of Fig.~\ref{fig:giniratio1} and Fig.~\ref{fig:giniratio2}. In the left panel, despite there exists a wide region in which the transition is not possible, we observe that if the fraction of agents considering themselves insulated from shocks is not too high or the initial fraction of the Green economy is not too low, we find fertile conditions for which the agents will move away from the Brown economy even in a no explicit policy setup. It is worth noticing that when the transition is possible, the iso-line for $T_2T$ suggests a form of exchangeability for these two dimensions. In the right (e) panel instead, the transition is not observed for any pair of these two parameters. Even if the fraction of the Green economy is close to 50\% and all agents take into account the externalities in their utility function, the extreme level of heterogeneity of the wealth distribution prevents any transition. This pinpoints that the level of heterogeneity of the wealth can play a critical role in inhibiting the transition in a range of values compatible with the one measured for the most developed countries, according to the scenarios provided by this model. 

\begin{table}[t!]
\caption{\label{tab:parameters2}State variables we observe.}
\begin{ruledtabular}
\begin{tabular}{lll}
\textbf{State variable} & \textbf{Notation} & \textbf{Unit}\\
\hline 
Total wealth of each asset & $W_\mathrm{G}$ or $W_\mathrm{B}$ & \$ \\
Returns of capital & $r_\mathrm{G}$ or $r_\mathrm{B}$ & \% \\
Time to Transition & $T_2T$ & years \\
Gini coefficient of wealths & $Gini$ & \% \\

\end{tabular}
\end{ruledtabular}
\end{table}

We now discuss other relevant interactions of the dynamics \textit{vis-à-vis} some selected parameters.

Focusing on panel (a) of Fig.~\ref{fig:giniratio1}, we observe -consistently with the model specifications - a high value $\lambda$ increases the relative weight of the externality awareness in the utility function (see Eq.~\eqref{eq:RHS}) and therefore facilitates the transition to a Green state. Similarly to the pair $\varphi_\mathrm{im}-ratio_\mathrm{G}$, the iso-lines for $T_2T$ suggest a sort of exchangeability mechanism for the initial heterogeneity of the system and the awareness $\lambda$. Namely, the model suggests that to preserve the transition time in a scenario with increasing wealth heterogeneity, the relative weight of the externality awareness in the utility function of the agents must increase. The relationship appears to be \textit{quasi}-linear in the upper left corner, the \textit{quasi}-linearity does not hold approaching the boundary line at which the transition does not take place anymore but the iso-lines stay monotonous.\\

The initial wealth and the fraction of agents considering themselves insulated $\varphi_\mathrm{im}$ appears to be \textit{de facto} unrelated for the dynamics of the system (see Fig.~\ref{fig:giniratio1}-d) and in this plane the transition time is essentially driven by the initial heterogeneity level. 

Consistently with the model specifications, a high $I$ generally tends to slow down or hinder the transition because it enhances the spread between $r_\mathrm{B}$ and $r_\mathrm{G}$ (see Fig.~\ref{fig:giniratio1}-g). The panel (g) interestingly shows an interplay of this spread with the initial level of heterogeneity (again a sort of interchangeability). The monotonicity of the iso-lines shows that the effects of an increasing wealth heterogeneity can be mitigated by reducing the spread of the yield of the two economies. The spread $I$ acts as a counteracting mechanism to the tendency of locking the system in the Brown state when the wealth heterogeneity increases. 

The impact of $r_\mathrm{loss}$ on the transition time is more complicated and not always monotonic (third column in Fig.~\ref{fig:giniratio1}). The interplay of this parameter exhibits different regimes because it counteracts the creation of externalities due to the Brown economy. Therefore its impact depends on whether the model is in a configuration locked in the Brown state where it will dynamically equilibrate the creation of new externalities by inducing new shocks, in a configuration where the transition is possible, or on the climate awareness level and the fraction of agents accounting for externalities in their utility function. The last two dimensions for instance can induce a dephasing for the agents \textit{vis-à-vis} of their \textit{ex ante} estimation of the costs due to the shocks in their utility function and the realized loss due to the level of Brown wealth. \\

In summary the possibility for the models in its base setup (i.e. the model without fiscal policies) to undergo the transition is strictly tied to the level of wealth heterogeneity, there is a level above which the transition is not possible anymore. Those levels are compatible with the ones observed in most developed countries. We observe the emergence of pair of parameters ($\lambda-Gini_{0}$, $\varphi_\mathrm{im}-ratio_\mathrm{G}$, $I-Gini_{0}$) whose interplay suggests mitigation axis in order to facilitate the transition time and to complement the policy recommendations we will discuss in Section~\ref{sec:policies}. 

We also want to stress that, so far, exchangeability mechanisms, dimension interplays and dynamics evolution have been discussed \textit{vis-à-vis} the transition time only (i.e. the end point of the simulations). In the next section, we are going to show that we can enrich the interpretation of the different scenarios by looking at the time dynamics along system-wide dimensions as the average growth of the systems, the rate of environmental destruction, and wealth inequalities. 

\subsection{Dynamics of some simulations and features at system scale}
\label{sec:dynamics_nopolicy}
Let us now look at the main features of the system in its baseline case for system-wide dimensions over time. We are going to consider the system initialized in its base set of parameters (see Table~\ref{tab:parameters}) for three levels of initial wealth heterogeneity, namely $Gini_{0}=0.7, 0.76, 0.85$. The three levels of heterogeneity correspond to three distinct regimes for our set of base parameters: i) for the first value we observe the transition to the Green economy well before $t_{max}$, ii) for the second value we are exactly at the boundary line between having and not having the transition, iii) for the third value, as argued in the previous section, the system is locked in the Brown economy. 

\begin{figure*}[!t]
    \includegraphics[width=1\textwidth]{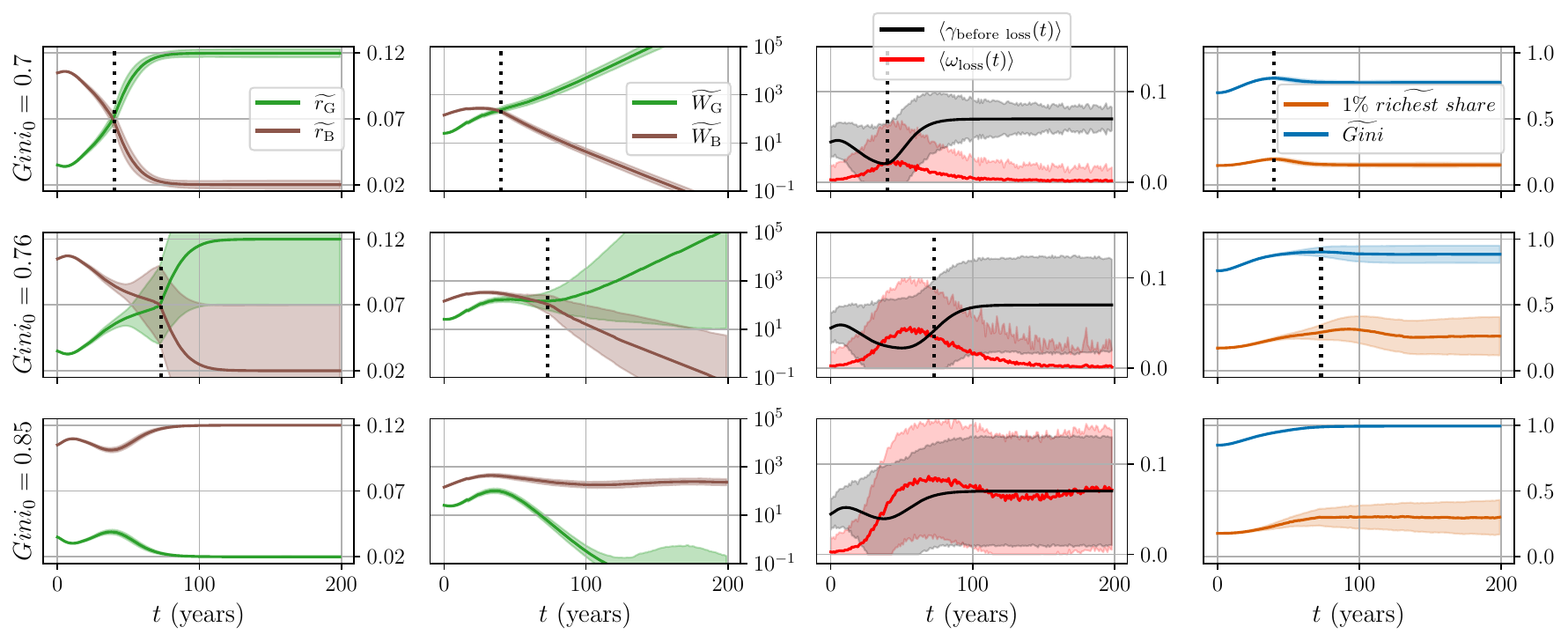}
    \caption{We plotted median trajectories for returns (1st column), median total wealths (2nd column), average share of loss $\omega_\mathrm{loss}$ and share of incomes $\gamma_\mathrm{before \ loss}$ on aggregate wealth  (3rd column), and median indicators for inequalities (4th column). All computations are made using reference parameters over 1000 Monte Carlo runs.}
    \label{fig:returns}
\end{figure*}

In the first column of Fig.~\ref{fig:returns} we illustrate the behavior of the evolution of the yield of the Brown and Green economy in time for the three regimes. The shaded areas represent the typical span of the ensemble of trajectories in order to average out the fluctuations due to the stochastic components of the simulations (i.e. the shock occurrences and their magnitude, both terms are set on average).  
As a general pattern of the model, simulations initially tend to drift towards an increase of the wealth in the Green Economy (NB: we have only two economies, therefore yield movements are mechanically symmetrical). A full transition (first panel) occurs only if the transfer of wealth from the Brown to the Green happens sufficiently fast. Whether this \textit{fast enough} threshold is satisfied depends on the balance between the revenue stream that can be invested into new Green wealth at each time step and the agent concern assign, in their utility function, to the externalities of the Brown economy. This balance is itself shaped by the initial level of inequality ($Gini_0$), since inequality shapes how wealth--and therefore investment capacity--is distributed across agents. Approaching the boundary level $Gini_0  \approx 0.76$, we observe that -- depending of the specific realization of the stochastic components of the model -- only about half of the simulations exhibits a transition. As the initial level of wealth heterogeneity increases, the shift toward the Green economy becomes progressively slower, eventually reaching a point at which the system remains frozen in its Brown phase. As discussed in the previous section, the fraction $\varphi_{im}$ of the wealthiest agents \textit{de facto} locks a large share of aggregate wealth into the Brown economy, leaving insufficient revenues available for investment in the Green economy and keeping the system too far from the relevant inflection region of the climate-shock probability function (see Eq.~\eqref{eq:pshock}).

Let us now explore how system-wide observables behave in these three scenarios over time. In particular we want to focus on the evolution of the total Brown and Green wealth (second column of Fig.~\ref{fig:returns}), the overall fraction of wealth destroyed by climate shocks (red line, third column of Fig.~\ref{fig:returns}), the overall growth rate of the economy before accounting for climate shocks (\textit{black} line, third column of Fig.~\ref{fig:returns}), the evolution of wealth heterogeneity in the fourth column of Fig.~\ref{fig:returns} (we track two indicators, the Gini index of the wealth distribution and the fraction of wealth owned by the top 1\% wealthiest agents.\footnote{The former indicator is more a mixed bulk/tail indicator for the heterogeneity while the latter is a purely tail driven indicator.}) 

We observe the gross growth rate of the wealth in the long run (\textit{black} line, third column of Fig.~\ref{fig:returns}) is similar regardless of the initial wealth heterogeneity or the occurrence/not occurrence of the transition. This is consistent with the fact that the gross yield of two economies is similar when one of the two economies represents the majority of the wealth available. What is dramatically different across the three scenarios is the net growth rate after accounting for the average fraction of wealth destroyed over time (red line, third column of Fig.~\ref{fig:returns}) and consequently the overall wealth evolution. If the transition occurs (first line and partly second line of Fig.~\ref{fig:returns}) the fraction of wealth destroyed by climate shocks after an initial peak will tend to zero and therefore the gross rate and the net rate will be \textit{de facto} the same. This is because the externalities due to the Brown economy will fade away in time when the wealth will become predominantly Green. On the other hand, if the transition does not occur, the fraction of destruction completely offsets the gross growth (on average) on the long run -- third panel, third column of Fig.~\ref{fig:returns} --  producing a stagnant system as illustrated by the flat brown line of the last panel of the second column of Fig.~\ref{fig:returns}. 
The wealth evolution in the non-dominant economy in all the three panels of the second column goes down because of the amortization term. 

In all three scenarios, agents are acting rationally according to their utility function. In other terms, they are behaving optimally from an individual point of view. The conclusion changes completely when we look the system at the aggregate scale. When the transition does not occur, the model produces scenarios which are suboptimal when they are benchmarked \textit{vis-à-vis} of the aggregate indicators as the net growth of the system. The wealthiest top 1\% of agents in the third scenario of Fig.~\ref{fig:returns} are marginally wealthier on a relative basis than the top 1\% of agents in the first scenario (where the transition occurs) as shown by the fourth column of this figure but they are significantly less wealthy on an absolute basis because the system growth is stagnant for non transition outcomes. This observation roots the need of exploring different types of fiscal policies which would allow to simultaneously have the agents to be optimal \textit{vis-à-vis} of their individual utility function and to produce outcome \textit{optimal} \textit{vis-à-vis} of aggregate dimensions such as the net growth rate and the fraction of destruction due to the externalities (e.g. between two scenarios with comparable time to transition, comparable net growth rate on the long run, we may prefer policies giving rise to less cumulated destruction or lower rate of destruction or requiring less re-distribution of wealth/revenues on a relative basis to achieve the same outcome). 

The evolution of the inequality across the three scenarios -- last column of Fig.~\ref{fig:returns} -- on one hand confirms that the level of heterogeneity can inhibit the transition but, interestingly, on the other hand the heterogeneity increases (or not decreases) when the transition occurs. The scenario with the transition is not \textit{naively} implying a less heterogeneous and less in-equal system on the long run. On the contrary, as we are going to see in the next section, in some cases the heterogeneity can contribute to speed up the transition when it is due to agents predominantly favoring allocation to Green. The model suggests that it is not the heterogeneity \textit{per se} which is bad or good for the transition, a system able to make the transition can stay or can even increase its initial wealth heterogeneity.  

\section{POLICIES: MODELING AND RESULTS}
\label{sec:policies}
In the baseline configuration we have seen that, although the agents are aware of the costs due to the externalities of the Brown economy, the transition does not occur for large regions of our parameters (or can take several decades). We have also observed that when the transition is slow or does not occur, the overall system is sub-optimal \textit{vis-à-vis} of system-wide dimensions as the net growth of the system and the cost for the system due to rate of destruction induced by the Brown economy. We therefore want to introduce a set of stylized and minimal fiscal policies in order to explore mechanisms of income re-distribution aiming at reconciling the mismatch between single agent optimization and system-wide benchmarks. These mechanisms can be: i) aware or unaware of the type of economy the agent wealth is invested in and/or ii) targeted—for instance through differentiated taxation or investment subsidies—or un-targeted.

The minimal and stylized mechanisms we will choose in the following will mimic two commonly observed features in a large fraction of countries: a progressive tax rate up to a certain income level, followed by a regressive rate above that level.  This regressive nature for the top income can be traced back to various factors: tax reduction originated by several elements, income deriving from capital gains that typically have significantly lower tax rate than the marginal tax rate of the regressive system, access to tax optimization scheme and instruments not available or not beneficial below specific thresholds, etc.

\subsection{Modeling of the policies}
Here we first discuss the progressive-regressive backbone of the stylized taxation mechanism and we then discuss the specifications of each policy.

\begin{figure}[t!]
    \includegraphics[width=0.6\columnwidth]{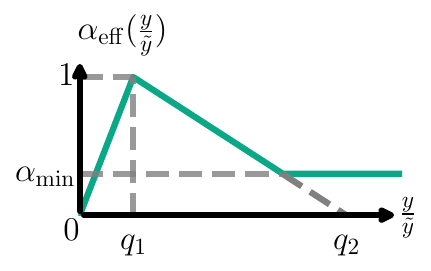}
    \caption{Effective factor for the multiplicative tax rate as a function of income $y$}
    \label{fig:effective_rate}
\end{figure}

In order to account for non stationarity of the total wealth (and consequently of the absolute income level), we index the tax mechanism to the median income at time $t$. We set two reference levels -- $q_1=20$ and $q_2=100$ times the median income -- which are pivotal in the progressive-regressive change of regimes for our taxation system (see Fig.~\ref{fig:effective_rate}). The taxation will be progressive up to 20 times the median income and regressive from this level up to 100 times the median income. It will then reach its minimum level and stay in a plateau. The first threshold has been chosen close to what has been empirically found in accordance for instance with Guzzardi et al. (see~\cite{guzzardi2024reconstructing} for Italy). For the second threshold, we have several evidences that the effective taxation is regressive for extremely high income/high wealth individuals, there are estimates of the typical effective tax rate for the top fortunes in some  countries (see Bach~\cite{bach2025billionaires}) but we are not aware of any empirical evidences for the level at which this effective rate is reached. We have therefore set it to an arbitrarily large value and the results do not depend significantly on the specific \textit{large} value we plug in the model. 

In details, we denote the median income as follows:
\begin{eqnarray}
    \tilde{y}(t)=\Big(r_\mathrm{G}(t)w_{\cdot,G}(t)+r_\mathrm{B}(t)w_{\cdot,B}(t)\Big)\Big|_{\rho_i=50\%}
\end{eqnarray}
and the effective tax rate of an agent at time $t$ will be defined by a function of the ratio of the agent income and the median income $\tilde{y}(t)$. We adopt a parsimonious piecewise linear function mapping the income ratio to the effective rate. We impose that an agent with zero income faces a 0\% tax rate, an agent with income $q_1\tilde{y}$ faces the maximum rate $r_\mathrm{tax}$, agents with income between $q_1\tilde{y}$ \& $q_2\tilde{y}$ have a linearly decreasing effective tax rate, and agents with income above $q_2\tilde{y}$ will have a constant rate set $r_\mathrm{tax}\alpha_\mathrm{min}<r_\mathrm{tax}$ (see Fig.~\ref{fig:effective_rate} for an illustration of the tax schedule). In formula, our mechanism reads:

\begin{eqnarray}
\label{eq:alpha_eff}
    \alpha_\mathrm{eff}(y)= \left\{ \begin{matrix}
\frac{y}{q_1\tilde{y}} & : \ y<q_1\cdot \tilde{y} \\
\max{\left[\alpha_\mathrm{min},\frac{y-q_2\tilde{y}}{(q1-q2)\tilde{y}})\right]} & : \ y \geq q_1\cdot \tilde{y}
\end{matrix} \right.
\end{eqnarray}
and the effective tax rate applied to an income $y$ is $r_\mathrm{tax}\,\alpha_\mathrm{eff}(y)$. The parameter $r_\mathrm{tax}$ sets the nominal maximum tax level and it is proportional to the total amount of tax revenues collected. $\alpha_\mathrm{min}$ sets the lower bound of the effective tax rate for very high income, and thus sets by how much the system deviates from a progressive policy. High $\alpha_\mathrm{min}$ implies that, in heterogeneous enough systems, the wealthiest agents face an effective tax rate closer to $r_\mathrm{tax}$. \\

Building on this mechanism, we consider four policies:
\begin{enumerate}[label=(\roman*)]
        \item \textit{Basic Income} (\textit{BI} hereinafter) a basic income, in which all agents are taxed and revenues are rebated equally;
        \item \textit{Tax Brown and redistribute à-la basic income} (\textit{TaxB-BI} hereinafter), a carbon tax on Brown income stream and redistributes revenues equally to all agents;
        \item \textit{Tax All, Credit Green} (\textit{TaxAll-CreditG} hereinafter), a Green incentive, where all agents are taxed, and  subsidies go to Green investors by increasing the effective return on their Green investments;
        \item \textit{Tax Brown, Credit Green} (\textit{TaxB-CreditG} hereinafter) a variant of the previous policy where only the income originated from wealth invested in the Brown economy is taxed to finance the subsidies to the Green economy.
\end{enumerate}

Let us now discuss how a) the amount of taxes collected, b) the amount redistributed and c) the agents' modeling of the different taxation mechanisms depends on the four policies. 
\\
\paragraph{Tax collected:} in formula, the total tax revenue collected at time $t$ is 
\begin{eqnarray}
    \label{eq:def_tax_amount}
    T(t)=r_\mathrm{tax}\sum_{i\in{\mathcal{A}_{taxed}(t)}}\alpha_\mathrm{eff}\left(\frac{y_i(t)}{\tilde{y}(t)} \right)y_i(t),
\end{eqnarray}
 where $\mathcal{A}_{taxed}(t)$ denotes the set of taxed agents. This set for \textit{BI} and \textit{TaxAll-CreditG} is the set of all agents while for \textit{TaxB-BI} and \textit{TaxB-CreditG} is restricted to the agents owning Brown assets.

 For incentive-targeted policies (those with CreditG) we need in this case to adjust the amount $T(t)$ in order to get the effective tax revenues. The model works typically in a regime where $W_\mathrm{G}(t)<W_\mathrm{B}(t)$ - when the inequality is reversed long enough the model makes the transition - therefore to avoid unrealistic large amount of subsidies distributed to Green agents we scale down the $T(t)$ by the factor $\alpha_{credit}(t)$ which accounts for the relative size of the income due to the two economies, namely:
\begin{eqnarray}
     \alpha_\mathrm{credit}(t) = \frac{1}{Y_\mathrm{tot}(t)}\sum_{i | \mathrm{G}}y_i(t)
 \end{eqnarray}
It follows that the effective amount of taxed due by an agent at time $t$ reads as
$\alpha_\mathrm{credit}(t)\alpha_\mathrm{eff}\left(\frac{y_i(t)}{\tilde{y}(t)} \right)y_i(t)r_\mathrm{tax}$. 
In such a way, we make the policy not depend trivially on the number of agents investing in the Green economy and on the relative size of the two economies. For instance, we avoid abnormally high  rate of subsidies per unit of Green wealth at the beginning of the simulations where typically $W_\mathrm{G}(t)\ll W_\mathrm{B}(t)$.
 \\
\paragraph{Tax revenue distribution}: for the policies \textit{BI} and \textit{TaxB-BI} the total tax revenues are redistributed as lump-sum to all agents, each agent receives the amount $T(t)/N$. 

In the incentive-targeted cases, the subsidies are not distributed as lump-sum but they are proportional to the income if the agent invests it in the Green Economy via a rate $r_\mathrm{boost}(t)$ defined as: 
\begin{eqnarray}
     r_\mathrm{boost}(t) = \frac{T(t)}{Y_\mathrm{tot}(t)}
\end{eqnarray}
The subsidy is acting as a extra yield for the Green economy.
The rate $r_\mathrm{boost}(t)$ is the average nominal taxation rate at time $t$. A way to interpret $r_\mathrm{boost}(t)$ is the rate as if the nominal collected tax revenues were to be distributed to all agents (ie. as if the total income $Y_\mathrm{tot}(t)$ were coming from Green wealth). In the limiting case where all agents were taxed at rate $r_{max}$ (i.e. no regressive/progressive tax mechanism) $r_\mathrm{boost}(t)$ would become simply equal to $r_\mathrm{tax}$ and $\alpha_\mathrm{credit}=1$.

It is worth noticing that the incentive-targeted policies, under the same nominal parameters, will redistribute lower amount of tax revenues according to these specifications.
\paragraph{Agents' modeling of taxation mechanism}: we work under the hypothesis that agents are aware of the different policies and the base case utility function defined in Eq.~\eqref{eq:utility_function} is updated as follows in the four cases:
\begin{enumerate}[label=(\roman*)]
        \item \textit{BI}: the utility function is not modified.\footnote{More precisely, the basic income policy does not change agents' behavior because the amount due in taxes and the amount they will receive back do not depend on the nature of their wealth (Green or Brown.) In this way we say that this type of policy is \textit{not targeted}.};
        \item \textit{TaxB-BI}: \begin{align}
            \Delta M_i=\left[r_\mathrm{G}-\left(r_\mathrm{B}-r_\mathrm{tax}\alpha_\mathrm{eff}\left(\frac{y_i}{\tilde{y}}\right)  \right)\right]\frac{y_i}{Y_\mathrm{tot}};
        \end{align}
        \label{eq:UF_TaxBBI}
        \item \textit{TaxAll-CreditG}:
        \begin{align}\Delta M_i=\left[(r_\mathrm{G}+r_\mathrm{boost})-r_\mathrm{B}\right]\frac{y_i}{Y_\mathrm{tot}}
        \end{align}
        \label{eq:UF_TaxAllCreditG}
        \item \textit{TaxB-CreditG}: 
        \begin{align}
            \begin{aligned}
                \Delta M_i=& \Bigg[(r_\mathrm{G}+r_\mathrm{boost}) \\
            &- \left(r_\mathrm{B}-r_\mathrm{tax}\alpha_\mathrm{credit}\alpha_\mathrm{eff}\left(\frac{y_i}{\tilde{y}}\right)  \right)\Bigg]\frac{y_i}{Y_\mathrm{tot}}.
            \end{aligned}
        \end{align}
        \label{eq:UF_TaxBCreditG}
\end{enumerate}

Agents use only past values up to time $t$ to evaluate the utility function which is leveraged to determine their choice at time $t+1$. In other words they do not try to forecast these values and they use the values from the previous time step.

\subsection{Case of a full Basic Income (\textit{BI}) policy}

\begin{figure*}[!h]
    \includegraphics[width=\textwidth]{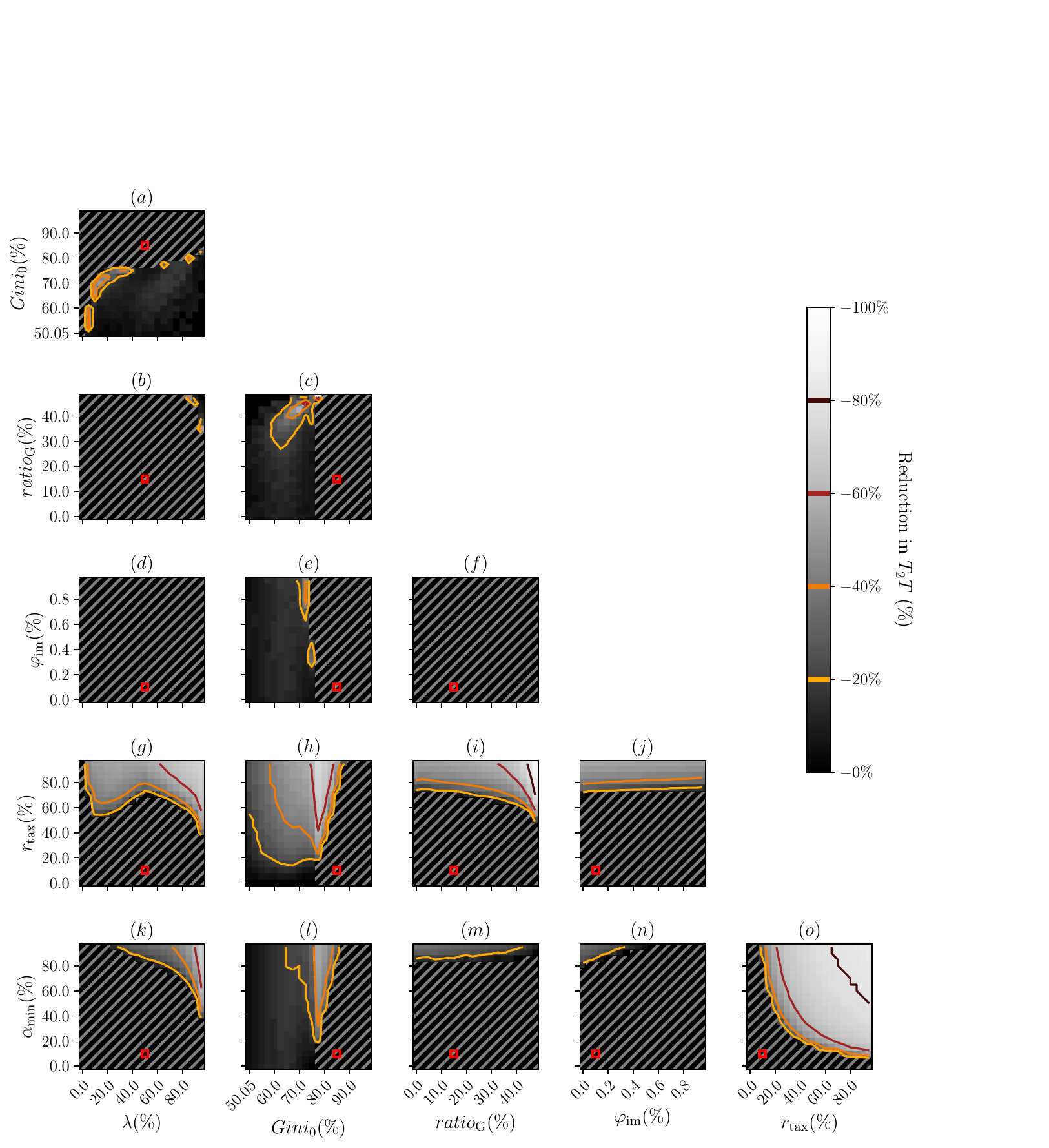}
    \caption{Phase diagrams of the relative reduction in transition 
    time $T_{2}T$ compared to the no-policy baseline, under a 
    non-targeted universal basic income (BI) scheme, with an initial 
    wealth inequality $Gini_0 = 0.85$ — consistent with empirical 
    estimates of wealth concentration in the United States — and a 
    maximum tax rate of $r_\mathrm{tax} = 10\%$. The hatched region 
    indicates parameter combinations for which at least half of the 
    Monte Carlo runs fail to transition within the simulation maximum time horizon.}
    \label{fig:BIpolicycolor}
\end{figure*}

We first consider the case of a non-targeted basic income scheme and examine how policy parameters affect the time to transition ($T_{2}T$). The change \textit{vis-à-vis} the base case discussed in Fig.~\ref{fig:giniratio1} -- the scenario with $Gini_0=0.85$ -- is reported in Fig.~\ref{fig:BIpolicycolor}. The basic income is \textit{de facto} reducing the heterogeneity level of the system and the role of income redistribution as an enabler for low-carbon transitions has been discussed in the literature (see~\cite{sumaila2024utilizing}). Our framework allows us to quantify this mechanism in a controlled setting. For each parameter combination, we compute the average percentage reduction in $T_{2}T$, relative to the no-policy baseline (base case).

Most regions in the phase diagrams appear dark for two distinct reasons. When the areas are hatched the systems fails to transition even with the activation of the \textit{BI} policy (NB: if the models do not undergo the transition under one of the policies, it cannot perform it in its base case too). In the dark non-hatched areas instead the system transitions in base case and the relative improvement for $T_2T$ due to the basic income is only marginal. We observe the largest benefits (lighter areas) from the basic income close to the transition/non-transition boundary because the \textit{BI} policy unlocks new areas for the transition where the system used not to transition in the no-policy setup (see Fig.~\ref{fig:BIpolicycolor} panel (a)). The effect also becomes significant in regions of high policy intensity (fourth and fifth lines of Fig.~\ref{fig:BIpolicycolor}): larger values of $r_\mathrm{tax}$ and/or $\alpha_\mathrm{min}$ mechanically increase the redistribution and therefore tend to accelerate the transition -- although the nominal maximum tax levels may sometimes be empirically unrealistic.

The basic mechanism underlying the effectiveness of this policy is that it reduces the mismatch between agents’ willingness to support the transition and their actual capacity to make it happen. A direct consequence is that narrowing this gap can shift capitals from the Brown economy to the Green economy more rapidly, thereby increasing the relative profitability of Green assets and making them more attractive to a broader set of agents.

Regarding the dependence on the initial $Gini_0$ (see Fig.~\ref{fig:BIpolicycolor}a, ~\ref{fig:BIpolicycolor}c, ~\ref{fig:BIpolicycolor}e), we observe the emergence of a critical threshold $Gini_0 \approx 0.8$, above which the transition is no longer achievable even under the \textit{BI} policy for the baseline parameter set. Let us now focus on Fig.~\ref{fig:BIpolicycolor}h and Fig.~\ref{fig:BIpolicycolor}l, which show that the policy parameters $r_\mathrm{tax}$ and $\alpha_\mathrm{min}$ both influence the location of this critical threshold. More precisely, the critical boundary in the $\alpha_\mathrm{min}$-initial $Gini_0$ phase diagram (Fig.~\ref{fig:BIpolicycolor}l) is similar to the critical boundary in the $r_\mathrm{tax}$-initial $Gini_0$ phase diagram (Fig.~\ref{fig:BIpolicycolor}h). This suggests that reducing the regressivity of tax mechanisms benefiting the wealthiest is as effective as simply increasing the overall tax rate. At the maximum tax rate $r_\mathrm{tax}$ or a zero-regressive tax mechanism ($\alpha_\mathrm{min}\rightarrow1)$, there remains a level of heterogeneity above which the transition cannot occur, around $Gini_0 \gtrsim 0.85$.

The emergence of a non-monotonic dependence on $\lambda$ in the $r_\mathrm{tax}$--$\lambda$ phase diagram (Fig.~\ref{fig:BIpolicycolor}g) suggests that the tax structure under a basic income system is not straightforward. Two mechanisms are indeed in competition: (i) when $\lambda$ is high, more agents invest in the Green economy at the first time steps, which facilitates the transition (already shown in Section~\ref{sec:phasediagram_nopolicy}); (ii) conversely, if several agents invest in the Green economy, income inequality is exacerbated, allowing the wealthiest investors taking advantage from the tax regressive mechanisms. Around $\lambda \sim 0.5$, the effect of the regressive component overcomes the benefit provided by increasing $\lambda$, leading to the emergence of non-monotonicity \\
Another interesting result comes from Fig.~\ref{fig:BIpolicycolor}o which shows the relationship between $r_\mathrm{tax}$ -- the maximum nominal tax rate applicable -- and $\alpha_\mathrm{min}$ -- characterizing the strength of the regressivity of the tax system.  Adjusting only one of the two parameters yields relatively little effectiveness. A combined action in both dimensions yields the best results.  In addition, we observe that there is a minimum value for both dimensions above which the system cannot make the transition: too high taxation in a highly regressive tax system is just as ineffective as a very low tax rate in a fully progressive tax system.\\

In summary, the model suggests that an untargeted policy can significantly affect the system near the transition/non-transition boundary, thereby enabling transitions in new regions. However, its effectiveness is ultimately limited by a strong initial level of inequality and by the endogenous feedback between the tax system, wealth concentration, and climate preferences. This limitation is further reinforced when the wealthiest agents have access to several ways of reducing the effective tax rate they face, bringing it well below the theoretical maximum applied to \textit{agile} taxpayers.

\subsection{Targeted policies}
\begin{figure*}[!h]
    \includegraphics[width=\textwidth]{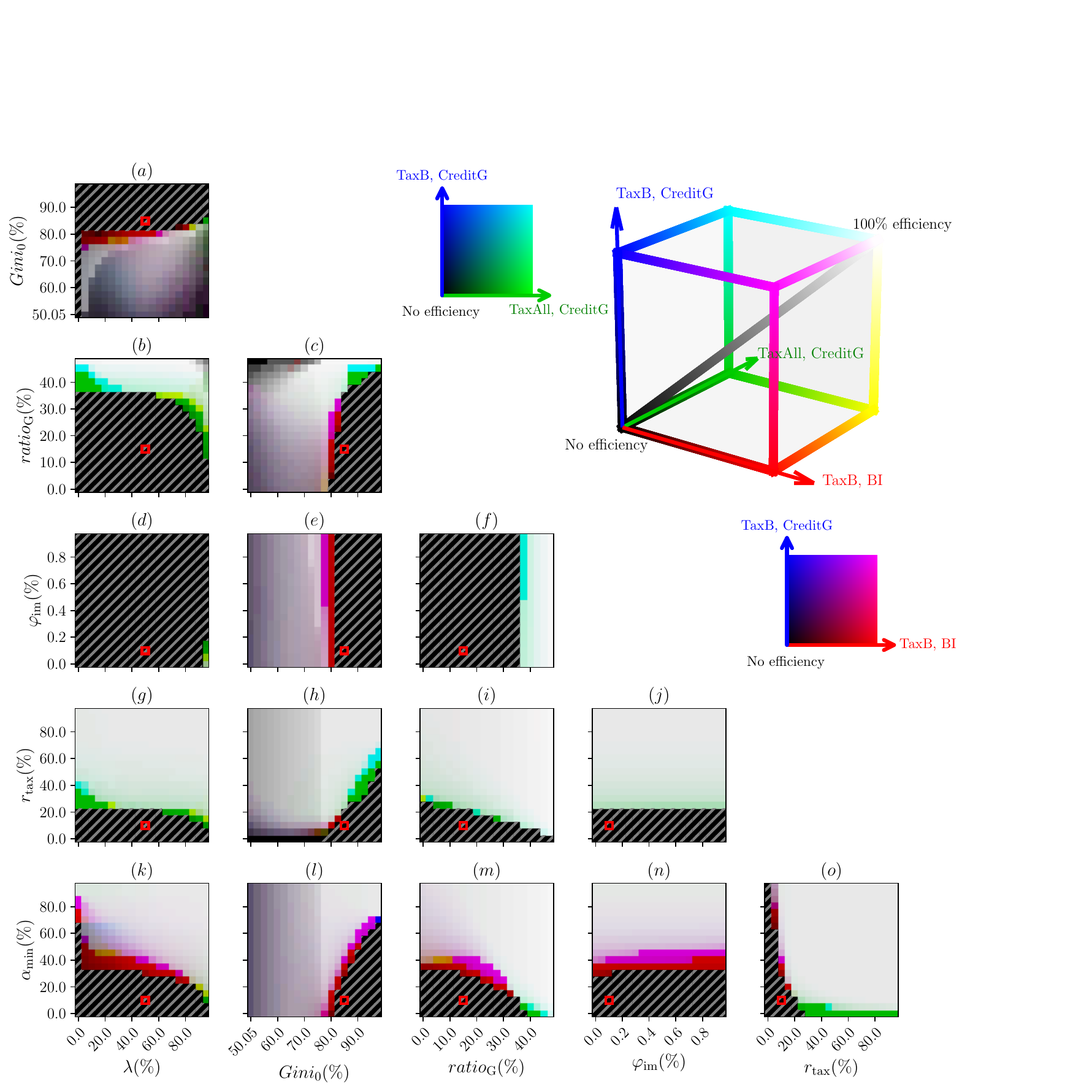}
    \caption{{Initial $Gini_0=0.85$} (except for panels \textit{a,c,e,h,l} where the dependence on the initial heterogeneity level is made varying) - Phase diagrams of the relative reduction in transition time $T_2T$ compared to the no-policy baseline, for the three targeted policies. The hatched region marks parameter combinations for which at least half of the Monte Carlo runs fail to transition (for the policy considered). \textit{red} (R), \textit{green} (G), and \textit{blue} (B) encode the efficiencies of the three policies; the displayed color at each point corresponds to their RGB combination. Lighter colors therefore indicate higher overall efficiency (larger reductions in $T_2T$).}
    \label{fig:allpoliciescolor1st}
\end{figure*}

\begin{figure*}[!h]
    \includegraphics[width=\textwidth]{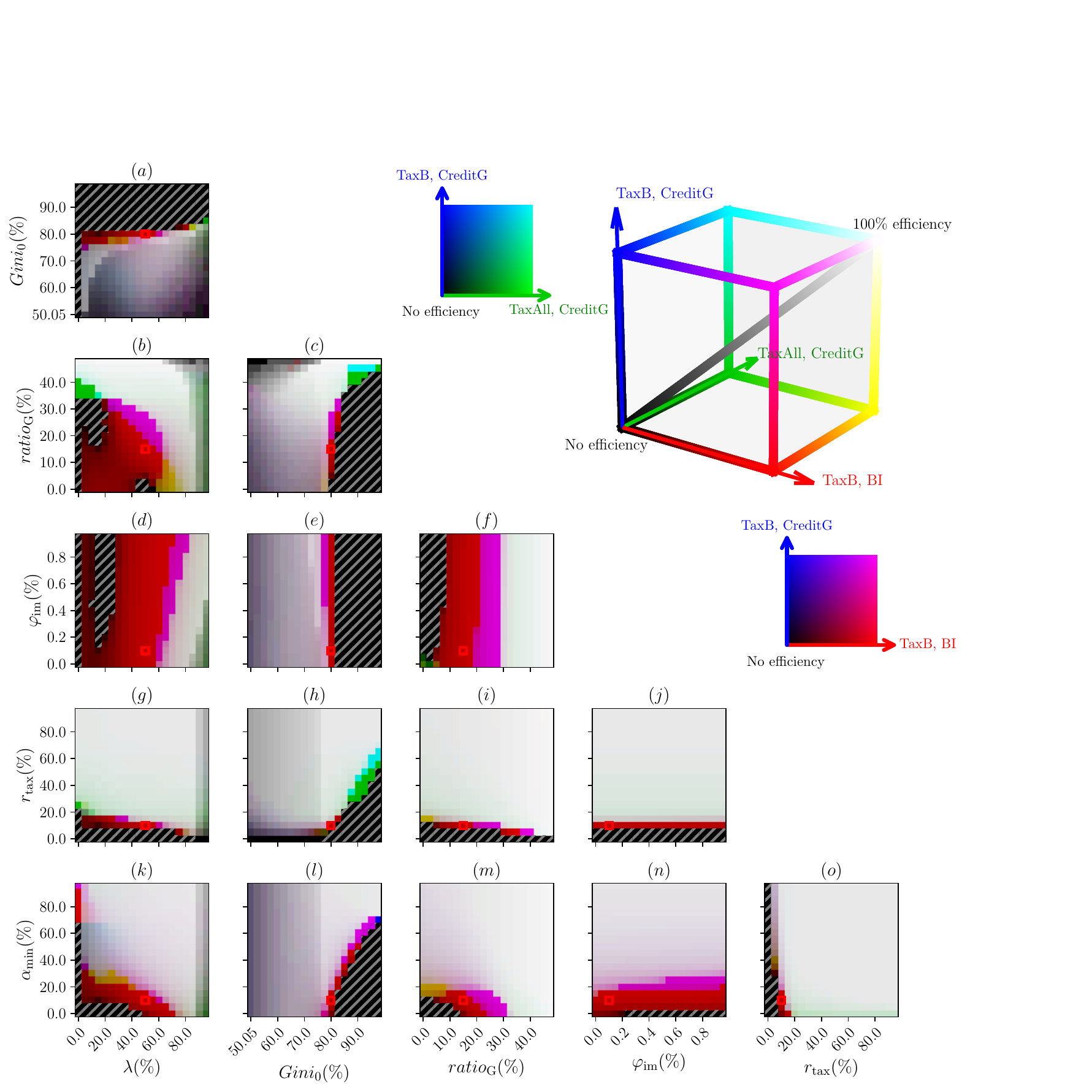}
    \caption{{Initial $Gini_0=0.8$} (except for panels \textit{a,c,e,h,l} where the dependence on the initial heterogeneity level is made varying) - Phase diagrams of the relative reduction in transition time $T_2T$ compared to the no-policy baseline, for the three targeted policies. The hatched region marks parameter combinations for which at least half of the Monte Carlo runs fail to transition (for the policy considered). \textit{red} (R), \textit{green} (G), and \textit{blue} (B) encode the efficiencies of the three policies; the displayed color at each point corresponds to their RGB combination. Lighter colors therefore indicate higher overall efficiency (larger reductions in $T_2T$).}
    \label{fig:allpoliciescolor2nd}
\end{figure*}
The three remaining policies are \textit{targeted}, i.e they   have an influence on agents' choices because the utility functions are modified. As done for the no-policy case, we first analyze the impact of the three targeted policies -- \textit{TaxB-BI}, \textit{TaxAll-CreditG} and \textit{TaxB-CreditG} -- \textit{vis-à-vis} the $T_2T$ and then we expand the analysis to other system-wide dimensions, such as the damages caused by climate-related events, economic growth, and the total tax revenues. \\

To quantify the effectiveness of the policies on $T_{2}T$, we compare each policy to the no-policy baseline by representing the reduction of $T_{2}T$ over the same set of pairs of parameters and initial conditions as done in the previous section for the \textit{BI} policy. We initially focus on the baseline set of parameters (see Tab.~\ref{tab:parameters}) and an initial $Gini_0=0.85$ -- this level of heterogeneity lies close to the transition boundary transition/non-transition (Fig.~\ref{fig:allpoliciescolor1st}). As for the basic income analysis, we compute the median reduction of the transition time (in percentage) due to the three targeted policies for each combination of the parameters. Differently from the \textit{BI} case, we have now three different values summarizing the impact of the three policies on $T_{2}T$ for each pair of parameter values. In order to make such a comparison, we encode the  reduction (\%) of $T_{2}T$ for each policy as a RGB color mix, by summing them in the RGB color space. In this representation, lightness indicates the overall effectiveness of the targeted policies (\textit{dark}: negligible  reduction of $T_{2}T$; \textit{bright}: significant reduction of $T_{2}T$), while saturation/different colors inform of  which policies have the largest impact in the different regions.\footnote{{Interpreting the phase-diagram colors:} In the scheme of Fig.~\ref{fig:allpoliciescolor1st} and Fig.~\ref{fig:allpoliciescolor2nd}, grayscale indicates that all three targeted policies have identical efficiency (\textit{black}: 0\%, \textit{white}: 100\%). The colors indicate that at least one policy outperforms the others; for example, \textit{red} means that \textit{TaxB-BI} policy is the most effective, while yellow indicates that both \textit{TaxB-BI} and \textit{TaxAll, Credit G} are equivalent \textit{vis-à-vis} $T_{2}T$ and are the most effective.}A scheme of the process is shown in Appendix~\ref{sec:appendix_RGB_construction}.

The darkest (\textit{near-black}) regions of Fig.~\ref{fig:allpoliciescolor1st} represent pair of parameters without transition: i.e. parameter sets for which we do not observe the transition even in presence of policies. In these cases, the policy efficiency is null and the regions are represented as hatched.
Interestingly, across all initial condition and parameter pairs, a substantial fraction of the phase diagrams appears in grayscale (i.e. \textit{white} / \textit{gray} tones). This means that in these regions, policies are beneficial but the details of the targeting mechanism are not critical. 

In panels (g-j, o) of Fig.~\ref{fig:allpoliciescolor1st} we show the dependence of the $T_{2}T$ reduction as a function of $r_\mathrm{tax}$ vs the other relevant dimensions. We observe the emergence of large \textit{light-gray} regions when $r_\mathrm{tax}$ increases: this is expected, higher tax rates generally increase policy effectiveness (larger reductions in $T_2T$). It is worth noticing that in the high (and likely unrealistic) tax rate regime the policy nuances of the targeting mechanism become indistinguishable (i.e \textit{white }color) and the only driver is \textit{de facto} the tax rate itself.

In panels (a,c,e,h,l) Fig.~\ref{fig:allpoliciescolor1st} we can explore the effectiveness of the policies for varying initial level of wealth heterogeneity.  In the regions where the transition already occurs without policies -- roughly speaking for initial $Gini_0 \lesssim 0.75$-- we observe tones of (dark) \textit{gray}. In our color scheme, this implies the policies typically reduce the $T_{2}T$ by up to a factor 2. As an example, in panel (a) of Fig.~\ref{fig:allpoliciescolor1st}, with $\lambda=0.8$ and $Gini_0=0.6$, the reduction in $T_{2}T$ is approximately 40\% for all the three targeted policies.\\

The regions with strong (and non-grayed) color saturation are the parameter regions where we observe the policies' highest effectiveness and where the effects of the target mechanisms are not equivalent for $T_{2}T$. These regions are typically located (see Fig.~\ref{fig:giniratio2}) where the system does not undergo the transition without the activation of the policy (for instance close to the  transition/non-transition boundaries in Fig.~\ref{fig:giniratio2}). In those regions, there exists a preferable policy, or at minimum, these results suggest several policy choices. For example, in panels (b, c) of Fig.~\ref{fig:allpoliciescolor1st}, for an initial $ratio_\mathrm{G}\approx 20\%$ close to the transition threshold, the policies \textit{TaxAll-CreditG} and \textit{TaxB-CreditG} achieve efficiency gains of about 88\% and 67\%, respectively, whereas \textit{TaxB-BI} has 0\% efficiency in this region. The results of panel (b) seem to indicate that there is an incentive to have targeted subsidies when the fraction of the Green economy is below 40\%. The panel (c), roughly speaking, shows that when inequality is high and $ratio_\mathrm{G}$ is high enough to get the transition, incentive targeted policies are preferable. As soon as the inequality is reduced and the share of the Green economy decreases, the policies targeted on the tax side (TaxB) emerge as the most effective.

This is a general pattern that we observe in the results: for high inequality the CreditG strategies are preferable while as soon as the inequality is lower the TaxB strategies emerge as more effective. We refer for instance to the comparison of Fig.~\ref{fig:allpoliciescolor1st} with Fig.~\ref{fig:allpoliciescolor2nd}. The latter summarizes the results with a lower initial $Gini_0=0.8$. A visual inspection reveals how now the \textit{magenta} / \textit{red} areas are much dominant than in the previous case (reminder: \textit{red} / \textit{magenta} correspond to the TaxB policies). There are several mechanisms at work that explain this behavior. When the system has a high degree of heterogeneity and a large enough fraction of Green economy in the system, a policy boosting the yield of the Green rather than reducing the inequality is preferable because of regressive mechanism. In such a regime of heterogeneity the wealthiest agents are more responsive to the yield of the economies rather than to the tax rate. On the contrary, when the system is less heterogeneous, the regressive mechanism is less at work and  targeted taxes turn to be the most effective strategy because now agents have a higher sensitivity to the rate in their choices. Roughly speaking, the results of the model suggest a subtle interplay among the level of heterogeneity, the share of Green economy and the level of regressivity of the tax mechanism. There exist scenarios where reducing the heterogeneity is not beneficial, because regardless of the nature of their wealth, the wealthiest agents might be predominantly driven by the yield rate of the Brown and Green economies, and therefore the transition would accelerate with policies targeting Green subsidies only. On the other hand, the more the level of inequality, the more targeted taxation systems should be favored. Our model suggests that defining optimal combinations of policies must leverage a sensitivity analysis in order to understand whether economic agents are more responsive to incentive or to taxation mechanisms, thereby maximizing the impact of any intervention.     

\begin{figure}[!h]
    \includegraphics[width=0.5\textwidth]{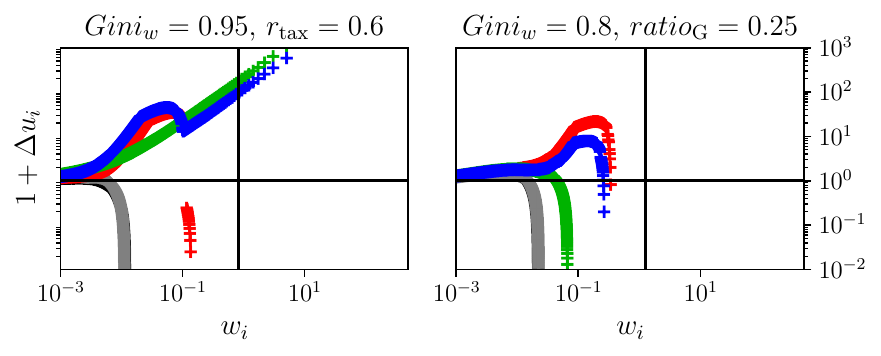}
    \caption{Average utility functions at $t=10\,\mathrm{years}$ for $10^3$ agents. The vertical line marks the wealth threshold defining the richest $\varphi_i$ agents, and the horizontal line indicates the strategy frontier: agents above it invest in Green at $t+1$, and those below invest in Brown. Colors follow the convention introduced in Fig.~\ref{fig:allpoliciescolor1st}. Each corresponds to a policy scheme: \textit{gray} for \textit{BI}; \textit{red} for \textit{TaxB-BI}; \textit{green} for \textit{TaxAll-CreditG} and \textit{blue} for \textit{TaxB-CreditG}.}
    \label{fig:policiesuf}
\end{figure}

\begin{figure}[!h]
    \includegraphics[width=0.5\textwidth]{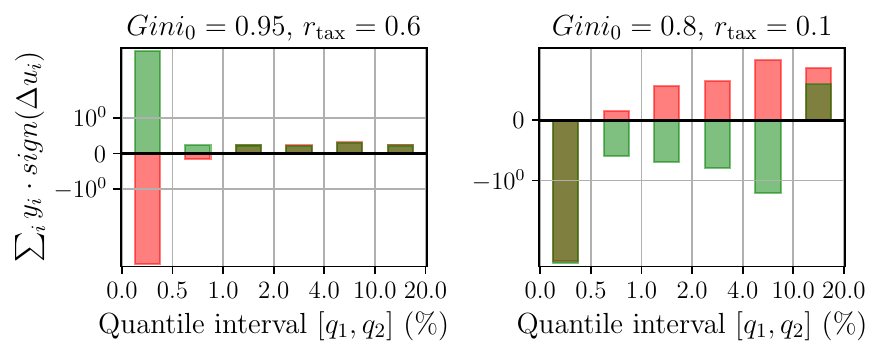}
    \caption{Sum of weighted $y_i$ at $t=10\,\mathrm{years}$ for different quantiles interval, and computed for $10^4$ agents. Red, is used for the simulation using policy \textit{TaxB-BI} and \textit{green} for the simulation using policy \textit{TaxAll-CreditG}. The left plot corresponds to a region where \textit{TaxAll-CreditG} is the most efficient, and the right plot is placed in a region where \textit{TaxB-BI} is the most efficient.}
    \label{fig:policiesyiuf}
\end{figure}

A way to visualize these types of effects in the model is to inspect the utility functions of the agents at a fixed time step. In the first plot of Fig.~\ref{fig:policiesuf} we report at $t=10$ years (i.e. after the first 10 time steps of the model) the difference in utility function $\Delta u_i$ between the two choices as a function of agent wealth. If $\Delta u_i>0$, Green economy will be chosen; and conversely $\Delta u_i<0$ will lead to a Brown investment.  In particular, the \textit{red} and \textit{blue} curves -- the curves for the tax-targeted policies TaxB -- display a pronounced peak corresponding to the region where agents face the maximum effective tax rate when choosing Brown, which strengthens the relative attractiveness of Green.\footnote{Only policies targeting taxes show that peak because in the two other cases (\textit{gray} curves for \textit{BI} and Green curves for \textit{TaxAll-CreditG} of Fig.~\ref{fig:policiesuf}), taxation levels do not depend on the nature of the investment and consequently they do not take into account the tax in their utility function.}In the left panel of Fig.~\ref{fig:policiesuf} inequality is extreme, and we see that only the \textit{blue} and \textit{green} lines (CreditG) enable the wealthiest agents to transition, consistently with the previous observation. When the inequality is reduced (right panel of Fig.~\ref{fig:policiesuf}) we see that TaxB emerges as the most effective mechanism. We want to stress that the \textit{blue} line also includes a CreditG component, but the transition is made possible by the TaxB component for two reasons: first, the \textit{green} line also sharing a CreditG component fails to promote the transition; and second, the \textit{blue} curves is above 0 for the wealthiest agents in its peak region, which we know is where $\alpha_\mathrm{eff}$ is maximum.

Fig.~\ref{fig:policiesyiuf} allows to see the relative impact of inequalities on the two policy mechanisms (TaxB vs CreditG): 
\begin{itemize}
    \item in the case of high inequalities, the left histogram in Fig.~\ref{fig:policiesyiuf} shows the cumulative signed incomes $y_i$ by quantile of wealth. The sign applied to the income is the previously defined condition $\Delta u_i>0$. At this level of heterogeneity, almost all the signed investment at stake is concentrated in the top quantile, and we see that this top quantile is investing in Green under a \textit{TaxAll-CreditG} policy, and in Brown under a \textit{TaxB-BI} policy;
    \item with a lower level of wealth inequality, the right histogram in Fig.~\ref{fig:policiesyiuf} shows that incomes are more evenly distributed among agents, and wealthiest agents are sensitive to the impact of the extra yield $r_\mathrm{boost}(t)$ (which is the same across agents), and are therefore more responsive to the impact of effective tax level. 
\end{itemize}

The model also produce scenarios where we observe a sort of over-targeting and inefficiency of the policy. This non-trivial pattern is marked by the presence of \textit{yellow}/\textit{orange} regions in Figs.~\ref{fig:allpoliciescolor2nd}. In our color combination scheme, it correspond to \textit{TaxB-CreditG} being the least efficient of the three targeted policies. This pattern can be seen, for example, in Figs.~\ref{fig:allpoliciescolor2nd}[b,c,i,k,m], where limited regions of the parameter space falls into this regime. As an illustration, for $ratio_\mathrm{G}=5\%$ and $\lambda = 0.7$ in Fig.~\ref{fig:allpoliciescolor2nd}b, the reduction in $T_2T$ is about 71\% for \textit{TaxB-BI}, 59\% for \textit{TaxAll-CreditG}, but null for \textit{TaxB-CreditG}.\footnote{We remind that all the percentage of reduction of $T_2T$ reported are the median value over the ensemble of simulations. The null reduction we observe for \textit{TaxB-CreditG} means that for at least half of the simulations we do not observe any reduction of $T_2T$.}

In regions characterized with a large initial gap $r_\mathrm{G}-r_\mathrm{B}$ (i.e. $ratio_\mathrm{G} \longrightarrow0$) and 
specific ranges for other parameters (e.g. $\lambda$ and $\alpha_\mathrm{min}$), targeting specifically the tax or the incentive side can have a significantly bigger impact on the utility function than targeting both at the same time.
Naively if we inspect the agent modeling of this policy --i.e. the effect of the policy on the utility function in Eqs.~[\eqref{eq:UF_TaxBBI}-\eqref{eq:UF_TaxBCreditG}]-- \textit{TaxB-CreditG} should be the most effective (or at least not the least effective). From this perspective, the existence of \textit{yellow}/\textit{orange} areas is puzzling. This phenomenon can be explained by going deeper into the analysis of the utility function. At first, the difference between \textit{TaxAll-CreditG} and \textit{TaxB-CreditG} is the level of subsidies, i.e. the boost $r_\mathrm{boost}(t)$ to the Green yield. The $r_\mathrm{boost}(t)$ can be significantly higher for the policy \textit{TaxAll-CreditG} rather than for the policy \textit{TaxB-CreditG} because $r_\mathrm{boost}(t)$ is directly proportional to the total tax revenue $T(t)$ (see Eq.~\eqref{eq:def_tax_amount}). This means that \textit{TaxAll-CreditG} can produce a higher spread of the yield between Green and Brown only via the boost of the Green yield than the double targeted policy \textit{TaxB-CreditG} for practically all agents (except for the higher taxed). 
When \textit{TaxB-BI} is considered, taxation is not reduced by the factor $\alpha_\mathrm{credit}$ due to the targeted distribution. As a result, the Green yield becomes more attractive to agents in this case rather than in \textit{TaxB-CreditG}, because of the stronger impact of taxes on the Brown yield. This again produces a spread between the two economies which is effectively higher than the one for \textit{TaxB-CreditG}.\\

In summary, the analysis of targeted policies reveals a subtle interplay between the level of wealth heterogeneity, the share of the Green economy, and the degree of regressivity of the tax mechanism. No single targeted policy performs best under every scenario. The model identifies distinct regimes in which one or two mechanisms become preferable. In particular, when inequality is high and a non-negligible fraction of the Green economy is already in place, incentive-targeted policies (CreditG) tend to be the most effective because the wealthiest agents —who own most of the investable income for the transition— are predominantly responsive to the spread between Green and Brown yields rather than to the effective tax rate they face. Conversely, when inequality is reduced, the regressive component of the tax mechanism plays a minor role and tax-targeted policies (TaxB) emerge as the most effective because agents become more sensitive to taxation in their utility function. We also identified non-trivial regimes —the \textit{yellow}/\textit{orange} regions in the phase diagrams— where the doubly-targeted policy \textit{TaxB-CreditG} is outperformed by the single-targeted counterparts due to the reduction of the Green-yield boost by the factor $\alpha_\mathrm{credit}$.

\subsection{Policy comparison beyond $T_2T$}

\begin{figure}
    \centering
    \includegraphics[width=\linewidth]{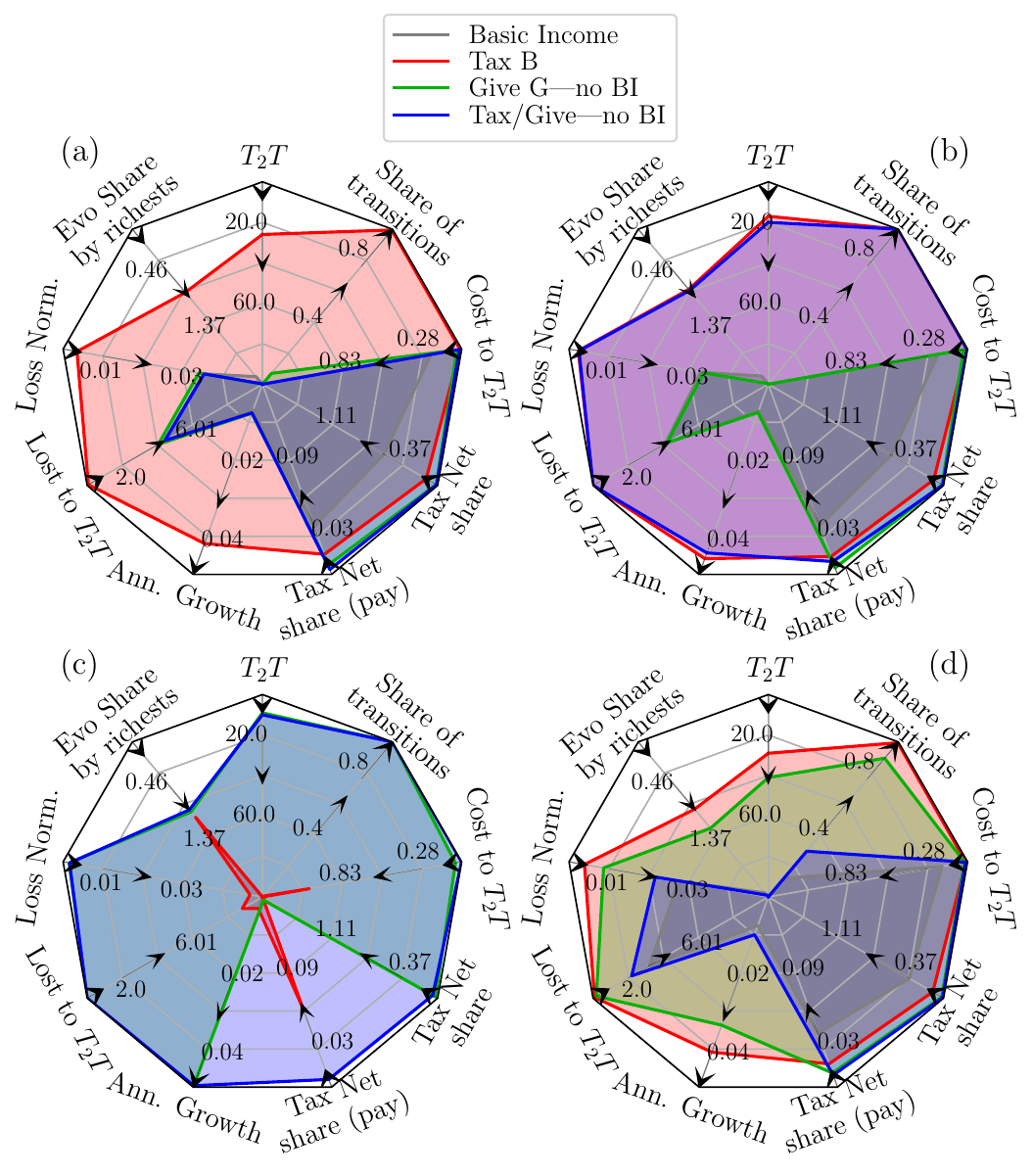}
    \caption{Comparison of policies efficiency for 9 different indicators, plotted in a way such that the biggest the area the better the indicator. We compute each radar plot based on parameters used in Fig.~\ref{fig:allpoliciescolor2nd}, in particular: (a) an initial $Gini_0=0.8$, (b) an initial $ratio_\mathrm{G}=25\%$, (c) an initial $Gini_0=0.95$ coupled with $r_\mathrm{tax}=60\%$ and, for (d) a $\lambda=0.7$ and an initial $ratio_\mathrm{G}=5\%$.}
    \label{fig:4radarplot}
\end{figure}

Similarly to the results without policies, we inspect the impact of the four policies beyond the $T_2T$ in order to fully characterize how they shape and change system-wide state. The success of a policy must obviously be benchmarked against its main target, but, at the same time, it is equally important to monitor and understand the impact on a broad range of dimensions. In particular, this can be a powerful approach when policies lie in regions where they have an equivalent effect on the reduction of $T_2T$. In Fig.~\ref{fig:4radarplot}, we summarize the results for the nine dimensions we consider. For each of them, the larger the surface on the radar plot, and the \textit{better} the policy. This means that a policy that completely covers the radar plot across all nine dimensions would be greatly beneficial simultaneously along all dimensions.\footnote{For most dimensions, the convention for defining \textit{better} is straightforward: it is better to have a quicker transition, a lower fraction of environmental destruction, a higher growth, the same effect considering lower tax revenues, etc. For some dimensions, namely those related to inequality, what is better is debatable and beyond the scope of this paper. We therefore adopt the convention that more equal systems are \textit{better}.}

Starting from the top axis of the radar plots, the first two indicators ($T_2T$ and \textit{Share of transitions}) represent, respectively, the median time to transition and the share of simulations in which a transition is effectively observed. Continuing clockwise, the following three indicators (\textit{Cost to $T_2T$}, \textit{Tax net share} and \textit{Tax net share (pay)}) are focused on the tax impact at an aggregate level for the agents: the first measures the share of cumulated incomes taxed relative to cumulated total wealth before the transition; the second gives the average share of each agent income that they received or paid under the policy, over the entire simulation period; and the last is defined like the second measure, but averaged only over agents who are net tax payers.\footnote{By net tax payers we mean agents for whom the raw amount of taxes minus the amount received through redistribution/subsides is positive.}The sixth indicator (\textit{Annualized Growth}) computes the annualized growth of the economy -- the \textit{CAGR} of the agents' total wealth. The seventh and eighth indicators measure the impact of climate shocks on the aggregate system. Namely, the first (\textit{Lost to $T_2T$}) measures the share of total lost wealth normalized by the total wealth before the transition. The second is the average share of loss in the economy over time (\textit{Loss Normalized}). The last indicator (\textit{Evo Share by richests}) tracks if and by how much the share of wealth owned by the $\varphi_\mathrm{im} \%$ richest agents has decreased ($<1$) or increased ($>1$) between $t=0$ and $t_\mathrm{max}$. More details on the nine indicators are provided in Appendix ~\ref{tab:radar_indicator}. \\

To illustrate these multidimensional trade-offs, we focus on four parameter combinations from Fig.~\ref{fig:allpoliciescolor2nd}, chosen as representative of four typical configurations: (a) the baseline case with initial $Gini_0 = 0.8$ (\textit{red} region), (b) the case with initial $ratio_\mathrm{G} = 25\%$ (\textit{magenta} region), (c) the case with $Gini_0 = 0.95$ and $r_\mathrm{tax} = 0.6$ (\textit{cyan}  region), and (d) the case with $\lambda = 0.7$ and initial $ratio_\mathrm{G} = 5\%$ (\textit{yellow} region). These four cases were selected because they capture qualitatively distinct policy regimes identified in Fig.~\ref{fig:allpoliciescolor2nd}, and therefore provide a compact but representative overview of the different mechanisms at play.
For each panel in Fig.~\ref{fig:4radarplot}, a clear pattern emerges: with the exception of the three tax-pressure indicators, the other indicators tend to evolve consistently across policies. This result supports the findings of Section~\ref{sec:dynamics_nopolicy}, where we showed that a faster transition is typically associated with more sustained growth and lower losses due to climate shocks.
That said, important differences between policies remain visible across the four radar plots. In particular, whenever the \textit{TaxB-BI} policy is in competition with other policies that also achieve the transition, it tends to perform better on most indicators, with the notable exception of the three tax-pressure dimensions (see Fig.~\ref{fig:4radarplot}(a,b,d)). We also observe that this policy performs best in reducing heterogeneities across all four cases considered, which is consistent with the fact that it is the only targeted policy whose redistribution includes a basic income component.

By contrast, \textit{TaxB-CreditG} appears to be the most tax-efficient policy across all four panels. Whether the transition occurs or not, this policy almost always performs slightly better on the three tax-pressure indicators. This can be explained by its highly targeted nature, which makes it comparatively more parsimonious in its fiscal intervention. Moreover, although the four panels suggest that \textit{TaxB-BI} is often more effective in promoting the transition, policies with a CreditG component still maintain relatively good outcomes across the full set of indicators, even when they fail to trigger the transition. Conversely, \textit{TaxB-BI} illustrates that a policy can be both very costly and highly ineffective when the conditions are unfavorable to its mechanism of action (see Fig.~\ref{fig:4radarplot}(c)).

Results show that policies leading to faster transitions generally also achieve stronger growth and lower climate damages. Among the policies, \textit{TaxB-BI} often performs best across most dimensions and is particularly effective at reducing inequality, although it tends to impose higher tax costs. In contrast, \textit{TaxB-CreditG} is the most tax-efficient policy due to its highly targeted design and consistently performs well on tax-related indicators. 

\subsection{Dynamics}
As done for the no-policy case, we want to study the time dynamics of some relevant dimensions in order to understand how the policies achieve their impact along the simulation. The variables we follow in time are the aggregate wealth allocation, the Green/Brown yields, the wealth heterogeneity , the level of environmental shocks and the fiscal pressure.

\begin{figure*}[t]
    \includegraphics[width=\textwidth]{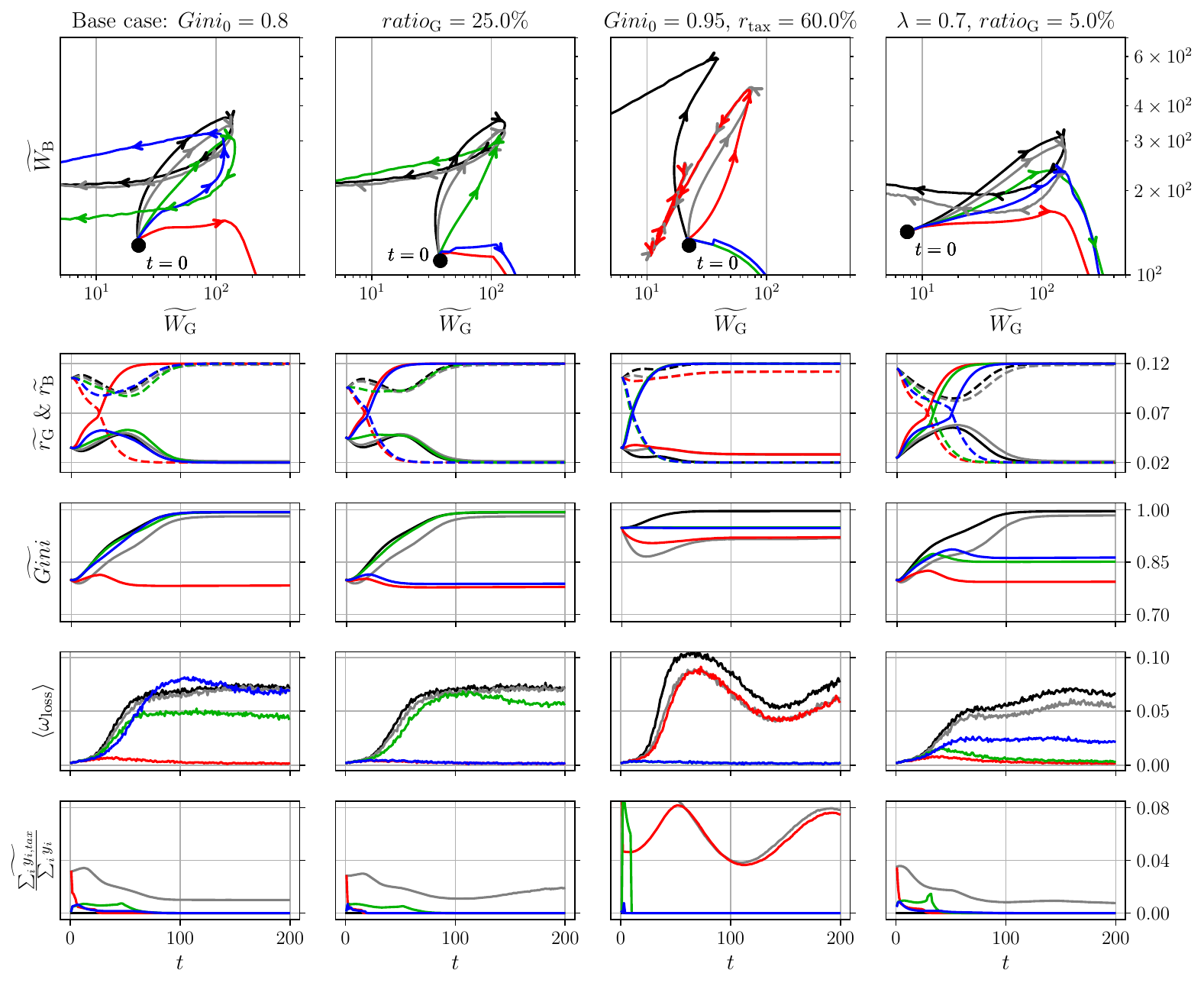}
    
    \caption{Median trajectories of (first row) the system's aggregate Green and Brown wealths; (second row) returns, with $r_\mathrm{G}$ shown as a solid line and $r_\mathrm{B}$ as a dashed line; (third row) the Gini coefficient; (fourth row) the loss share; and (fifth row) the share of gross taxes. The last row shows the system’s phase portrait. Colors denote policy regimes: \textit{black}, no-policy; \textit{gray}, basic income; \textit{red}, carbon tax (taxing Brown incomes with lump-sum rebates); \textit{green}, Green incentives (subsidies to Green investment); and \textit{blue}, the combined policy (carbon tax plus Green incentives). Simulations use the baseline parameter set (Table~\ref{tab:parameters}) with $10^3$ agents; trajectories are averaged over 2000 Monte Carlo runs.}
    \label{fig:dynamics_policies}
\end{figure*}

Figure~\ref{fig:dynamics_policies} shows the median (or the average depending on the specific indicators represented) trajectories over 2000 realizations of the simulations for the four policies + no-policy scenarios with the following color code: no-policy \textit{(black)}, basic income \textit{(gray)}, \textit{TaxB-BI} \textit{(red)}, \textit{TaxAll-CreditG} \textit{(green)} and \textit{TaxB-CreditG} \textit{(blue)}. We focus on the baseline parameter configuration with initial $Gini_0 = 0.8$, and take the four specific cases we studied in Fig.~\ref{fig:4radarplot}.

The first row of Fig.~\ref{fig:dynamics_policies} shows the system trajectories in the plane specified by the aggregate Green and Brown   wealth. In the first time steps of the simulations, all five cases exhibit an increase in both Green and Brown wealth. Beyond this initial phase, we identify three main regimes in the model. If the transition occurs, Green wealth grows indefinitely while Brown wealth declines toward zero. If, instead, the inversion of the yields does not occur, some trajectories instead converge to a regime characterized by stagnant Brown wealth and vanishing Green wealth due to recurrent climate shocks. 

In this last situation a third regime can be identified (third column of Fig.~\ref{fig:dynamics_policies}): an asymptotic pattern is observed under non-incentive-targeted policies—namely, \textit{BI} and \textit{TaxB-BI}. In these cases of high heterogeneity and raw taxation level, the system also converges to a stable in time Brown wealth level, but Green wealth is also tending towards a non-zero value. This outcome reflects the fact that non-incentive-targeted policies redistribute more wealth toward poorest agents (more invested in Green) than incentive-targeted schemes do.

The second row of Fig.~\ref{fig:dynamics_policies} shows the evolution of the two yields, $r_\mathrm{G}(t)$ is plotted with solid lines and $r_\mathrm{B}(t)$ with dashed lines. In the no-policy baseline, yields follow a characteristic pattern: during an initial phase lasting roughly 40--60 years, $r_\mathrm{G}$ gradually increases while $r_\mathrm{B}$ decreases, reflecting the early reallocation of wealth toward Green assets. However, this initial trend is not enough to reach the tipping point of the transition. Without policies supporting this initial virtuous dynamics the system frequently ends trapped in a regime in which $r_\mathrm{B}$ remains competitive, preventing the full transition.
It is worth noticing that in most cases, even when the policies will enable the transition, it takes a significant time to invert the spread $r_\mathrm{G}-r_\mathrm{B}$. Only when a sufficiently large amount of capital has been allocated to the Green-economy, the policies contribute to accelerate the transition.

In the third and fourth rows of Fig.~\ref{fig:dynamics_policies} we show the evolution of the wealth inequality as measured by $Gini(t)$ and the impact of the climate related destruction as measured by the share of losses $\omega_\mathrm{loss}(t)$. As already observed in the no-policy cases, both indicators display a strong dependence with the transition outcomes: a fast transition is generally associated to lower inequality and more limited climate related destruction. In particular, when \textit{TaxB-BI} succeeds in inducing a transition, it also tends to minimize the growth of inequality and climate-related losses. It is worth noticing, that achieving the transition typically also implies the reduction of the inequality except in the case of \textit{TaxB-CreditG} and \textit{TaxAll-CreditG} as previously mentioned (see \textit{blue} and \textit{green} lines in the third row of Fig.~\ref{fig:dynamics_policies}).
In the scenarios without a transition, the introduction of any policy reduces the average level of inequality and the average impact of climate shocks relative to the no-policy baseline even if policies do not achieve their primary goal.

The fifth row of Fig.~\ref{fig:dynamics_policies} provides a proxy for fiscal pressure by tracking the ratio of gross tax revenues to total income in the economy at each time step. As previously observed, the basic income policy is the one characterized by the largest fiscal pressure and, to some extent, its relative lower efficiency (i.e. given the same size of the total economy of the systems it requires more tax revenues to achieve the same transition than the targeted policies). When the initial inequality is high (see the third column of Fig.~\ref{fig:dynamics_policies}), \textit{TaxB-BI} becomes less effective, and most agents continue to invest in Brown. In this regime, \textit{TaxB-BI} has an overall effect very close to the behavior of the \textit{BI} policy. Given the initialization of our systems with $W_\mathrm{G}<W_\mathrm{B}$ and heterogeneity such that the regressive part of the tax mechanism is hit, all policies exhibit a peak in fiscal pressure during the first time steps, followed by a gradual decline. The most \textit{parsimonious} policy -- again in relative terms -- turns out to be \textit{TaxB-CreditG}.

\section{conclusion}
\label{sec:conclusion}

We have developed a stylized agent-based model to investigate how wealth heterogeneity may affect the dynamics of an ecological transition. Agents allocate their wealth between Brown and Green alternatives, facing a trade-off between short-term gains and perceived exposure to environmental damages. A central assumption of the model is that the sensitivity to such damages in the utility function decreases with wealth, which in turn reduces the aggregate resources effectively available for the transition. The simulations show that this asymmetry often generates an obstacle to the transition. In the absence of policy intervention, the initial degree of wealth inequality plays a decisive role: beyond certain levels, the economy remains locked in a Brown regime even when environmental damages are present and even when part of the population is sensitive to them.

The phase diagrams further show that transition outcomes depend on the interaction between inequality, initial Green capital, environmental awareness, perceived immunity from shocks, and the relative return advantage of the Brown sector. While successful transitions may imply lower short-term growth, they ultimately lead to more resilient long-run trajectories by reducing recurrent climate-related losses. Importantly, the model identifies critical levels of heterogeneity beyond which no transition occurs, regardless of the initial share of Green wealth in the system or of the policy implemented: in such regimes, the structural concentration of wealth dominates all other mechanisms and locks the system in its Brown state.

Below these critical levels, however, the role of inequality turns out to be far less univocal than one might naively expect. Policy experiments indicate that redistributive and targeted mechanisms can substantially expand the regions of parameter space in which transition becomes possible, but their effectiveness depends strongly on the initial distribution of wealth and on the degree of regressivity embedded in the fiscal mechanism. We even identify non-trivial regimes in which heterogeneity can be \textit{helpful}: for instance when the wealthiest agents become predominantly responsive to the relative yield of the two economies, so that incentive-targeted Green subsidies can leverage their concentrated investment capacity to accelerate the transition. Conversely, in less heterogeneous configurations, targeted taxation of Brown activities tends to be more effective, as agents are more sensitive to the effective tax rate they face. Untargeted transfers, in turn, are typically most useful near the transition boundary. These results emphasize that there is no universally optimal policy: it is essential to carefully identify the state of the system and the dominant behavioral drivers of its agents, in order to apply the appropriate type of intervention

Several extensions of the present framework appear particularly interesting. A first direction would be to introduce network effects, imitation, or social influence mechanisms, so that agents' investment choices depend not only on returns and perceived damages, but also on the behavior of neighboring agents or reference groups—either through herding at the micro level or through groups of agents aggregated into countries at the macro level. A second extension would be to allow for technological progress, endogenous changes in Green productivity, or policy-induced learning curves, which could shift the transition threshold over time and possibly unlock regimes that are currently locked in the Brown state. Finally, a third natural step, already alluded to in Section~\ref{sec:main_model}, is to relax the assumption of full rationality and equip agents with bounded rationality through, e.g., a logit decision rule with finite intensity of choice $\beta$. We leave these promising avenues for future work.

\section{Acknowledgments}
\label{sec:discussion}

We would like to thank Olivier Benhamou, Jean-Philippe Bouchaud,  Ortensia Forni, Alain Granjean, Elia Moretti for fruitful discussions. 
This research was conducted within the Econophysics
\& Complex Systems Research Chair, under the aegis of
the Fondation du Risque, the Ecole polytechnique and Capital Fund
Management.

\clearpage
\bibliography{bibfile}

\onecolumngrid
\appendix
\label{sec:appendix}

\clearpage

\section{Table of useful indicators used}
\label{sec:radar_indicator_appendix}

\begin{table*}[h!]
\caption{\label{tab:radar_indicator}Indicators used in the radar plot. We define by $Y_\mathrm{tax}(t)=\sum_iy_\mathrm{tax}(t)$ the raw total quantity of taxes took from agents by the policy, before any distribution. We also set $y_{i,\mathrm{nettax}}$ the net amount of incomes changed by the policy for agent $i$, i.e given by $y_{i,\mathrm{nettax}}(t)=y_{i, bonus}(t)-y_{i, tax}(t)$, then taking into account distributions (what agents receive).}
\begin{ruledtabular}
\begin{tabular}{lll}
\textbf{Name}  & \textbf{Description}\\
\hline 
$T_2T$ & Time such that $r_\mathrm{G}(t)>r_\mathrm{B}(t)$  \\
Share of transitions & Fraction of runs for which $T_2T<t_\mathrm{max}$  \\
Cost to $T_2T$ & $\frac{\sum_{t<T_2T}Y_{tax}(t)}{\sum_{t<T_2T}W_\mathrm{tot}(t)}$ \\
Tax net share & $\frac{1}{t_\mathrm{max}}\sum_t \left(\frac{1}{N} \sum_i \frac{y_{i,nettax}(t)}{y_i(t)}  \right)$  \\
Tax net share (pay) & $\frac{1}{t_\mathrm{max}}\sum_t \left(\frac{1}{N_\mathrm{net payers}} \sum_{i|y_{i,nettax}(t)<0} \frac{y_{i,nettax}(t)}{y_i(t)}  \right)$ \\
 Ann. Growth & $\left(\frac{W_\mathrm{tot}(t=t_\mathrm{max})}{W_\mathrm{tot}(t=0)}  \right)^\frac{1}{t_\mathrm{max}}-1$ \\
Lost to $T_2T$ & $\frac{\sum_{t<T_2T}W_\mathrm{loss}(t)}{W_\mathrm{tot}(T_2T)}$ \\
Lost norm. & $\frac{1}{t_\mathrm{max}}\sum_t \frac{W_\mathrm{lost}(t)}{W_\mathrm{tot}(t)}$ \\
Evo. Share by richests & $\frac{\sum_{i \in \varphi_\mathrm{im} rich} w_i(t=t_\mathrm{tax})}{W_\mathrm{tot}(t=t_\mathrm{tax})} \big/ \frac{\sum_{i \in \varphi_\mathrm{im} rich} w_i(t=0)}{W_\mathrm{tot}(t=0)}$ \\
\end{tabular}
\end{ruledtabular}
\end{table*}

\section{Construction of the RGB phase diagrams}
\label{sec:appendix_RGB_construction}
\begin{figure}[h!]
    \centering
    \includegraphics[width=0.35\linewidth]{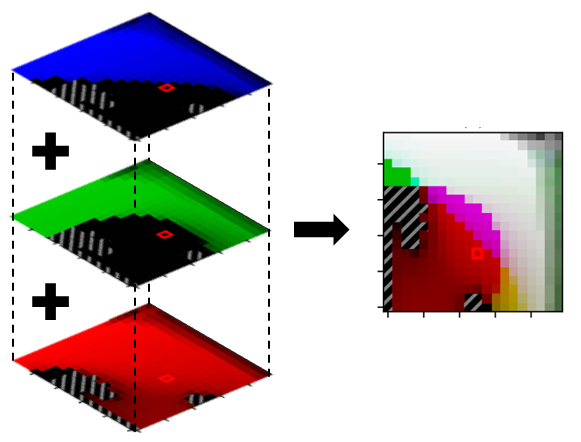}
    \caption{Scheme of the construction of the panel (b) in Fig.~\ref{fig:allpoliciescolor2nd}. For each policy we compute the effectiveness of each policy (the percentage of reduction in $T_2T$ compared to the no-policy case), and then we sum the 3 values obtained in the RGB space.}
    \label{fig:placeholder}
\end{figure}

\clearpage
\section{Initial Wealths}
\label{sec:initial_wealths}
\begin{table*}[h!]
\caption{\label{tab:gini_share} Share of wealth held by a fraction of the richest, for a given Gini. Generated with a Pareto Type-II distribution.}
\begin{tabular}{lllllllllll}
\multicolumn{1}{l}{}                  & \multicolumn{10}{c}{\textbf{$\mathbf{Gini_0}$ coefficient}}                                                                                                                          \\ 
\multicolumn{1}{l||}{\textbf{Fraction}} & \textbf{0.525} & \textbf{0.575} & \textbf{0.625} & \textbf{0.675} & \textbf{0.725} & \textbf{0.775} & \textbf{0.825} & \textbf{0.875} & \textbf{0.925} & \textbf{0.975} \\ \cline{2-11} \noalign{\vskip 0.85mm} \cline{2-11}
\multicolumn{1}{l||}{\textbf{1\%}}      & 6.78           & 9.91           & 14.3           & 20.1           & 27.5           & 36.6           & 47.5           & 60.3           & 74.8           & 91.2           \\
\multicolumn{1}{l||}{\textbf{0.1\%}}    & 1.18           & 2.04           & 3.81           & 6.84           & 11.7           & 18.9           & 29.3           & 43.5           & 62.1           & 86.0           \\
\multicolumn{1}{l||}{\textbf{0.01\%}}   & 0.15           & 0.39           & 0.98           & 2.28           & 4.90           & 9.71           & 18.0           & 31.3           & 51.6           & 81.0           \\
\multicolumn{1}{l||}{\textbf{0.001\%}}  & 0.02           & 0.07           & 0.25           & 0.75           & 2.04           & 4.98           & 11.0           & 22.5           & 42.8           & 76.4           \\
\multicolumn{1}{l||}{\textbf{0.0001\%}}                      & 0              & 0.01           & 0.06           & 0.25           & 0.85           & 2.55           & 6.77           & 16.21          & 35.50          & 72.0          
\end{tabular}
\end{table*}

\begin{figure*}

{\includegraphics[width=2.2in]{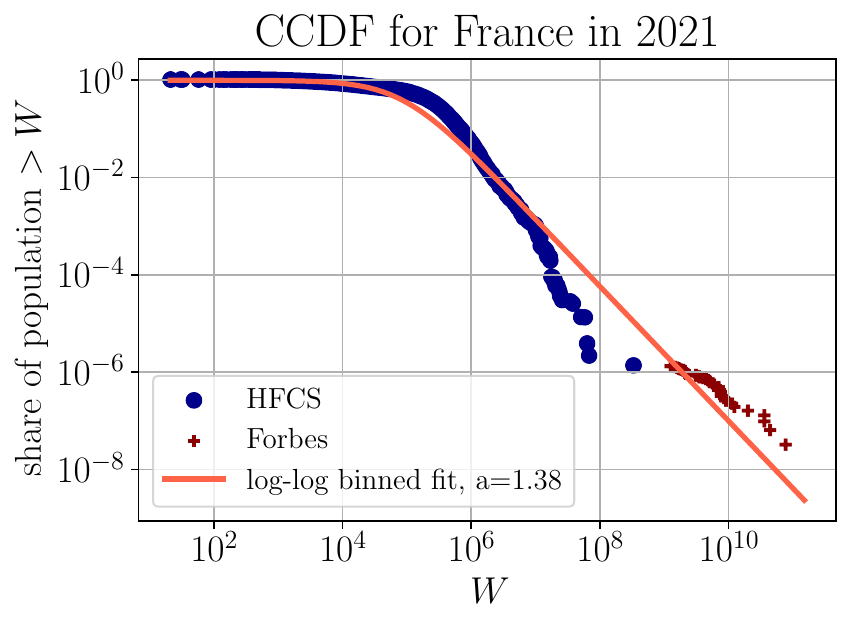}}%
\quad
 {\includegraphics[width=2.2in]{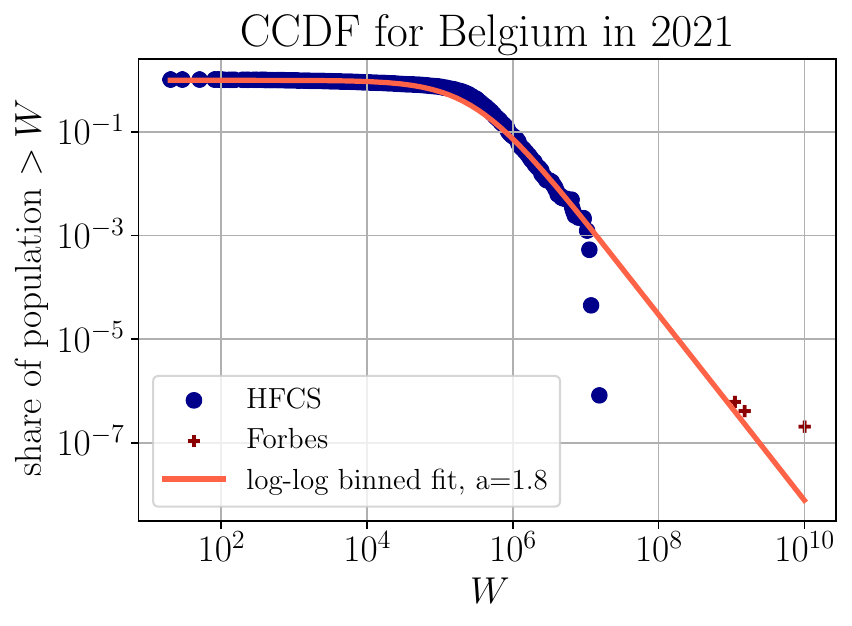}}%
 \quad
 {\includegraphics[width=2.2in]{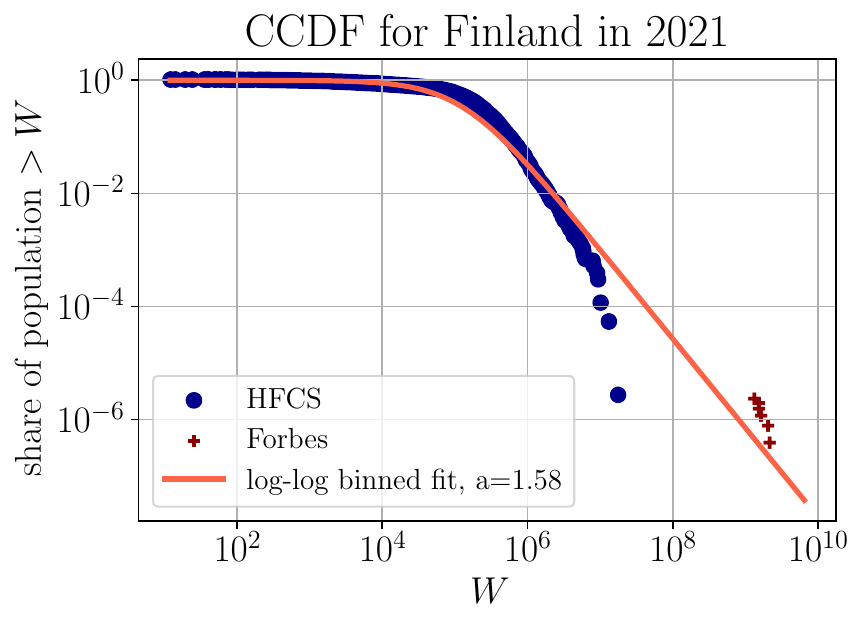}}%
 \quad
 {\includegraphics[width=2.2in]{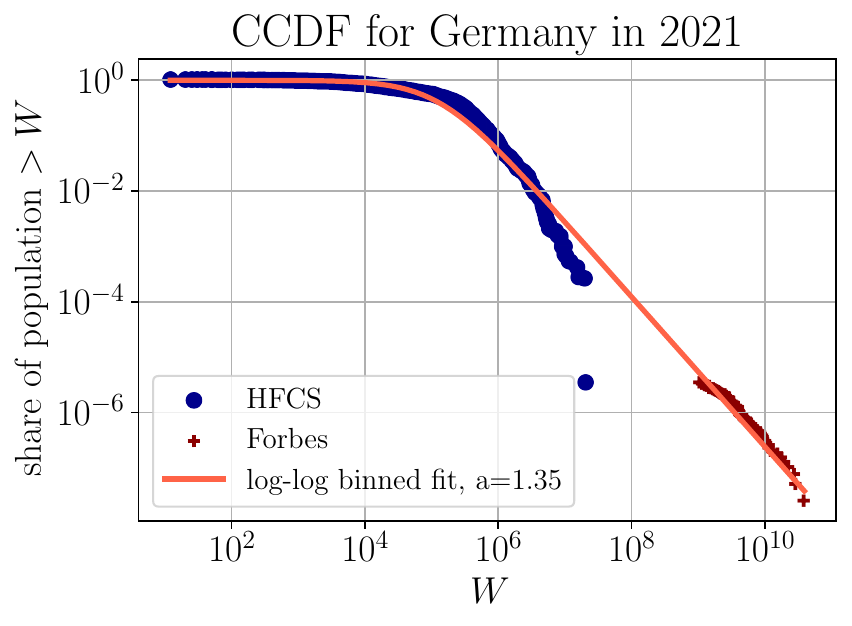}}%
 \quad
 {\includegraphics[width=2.2in]{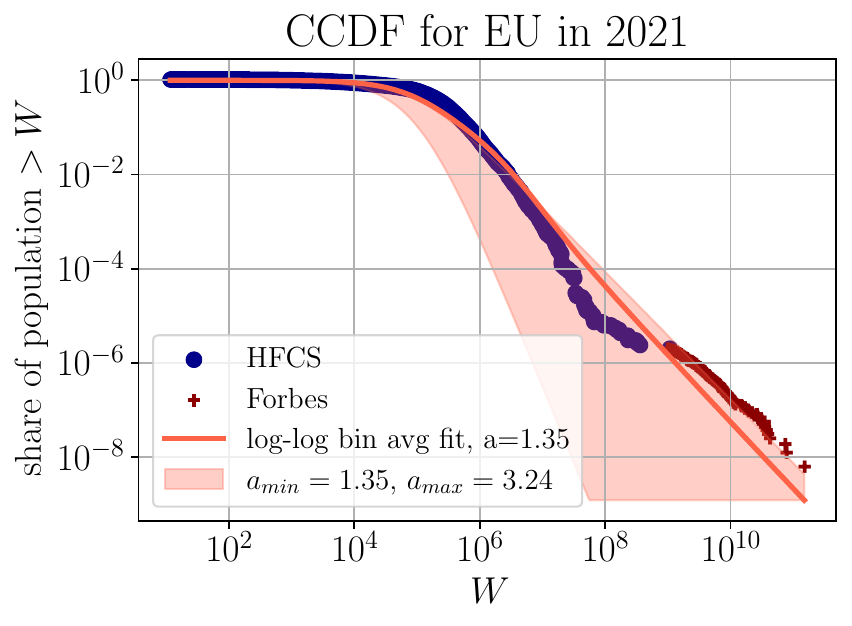}}%
 \quad
  {\includegraphics[width=2.2in]{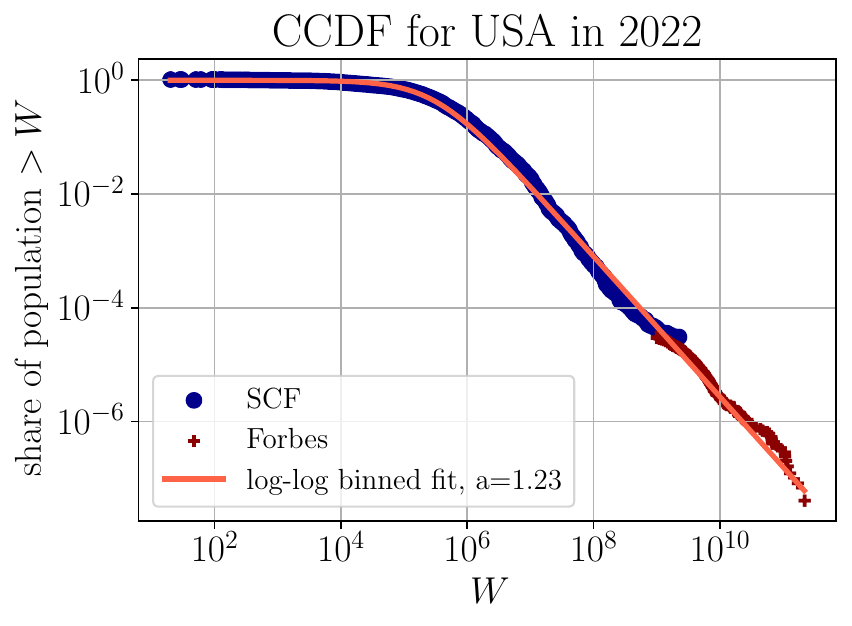}}%
 \quad
  {\includegraphics[width=2.2in]{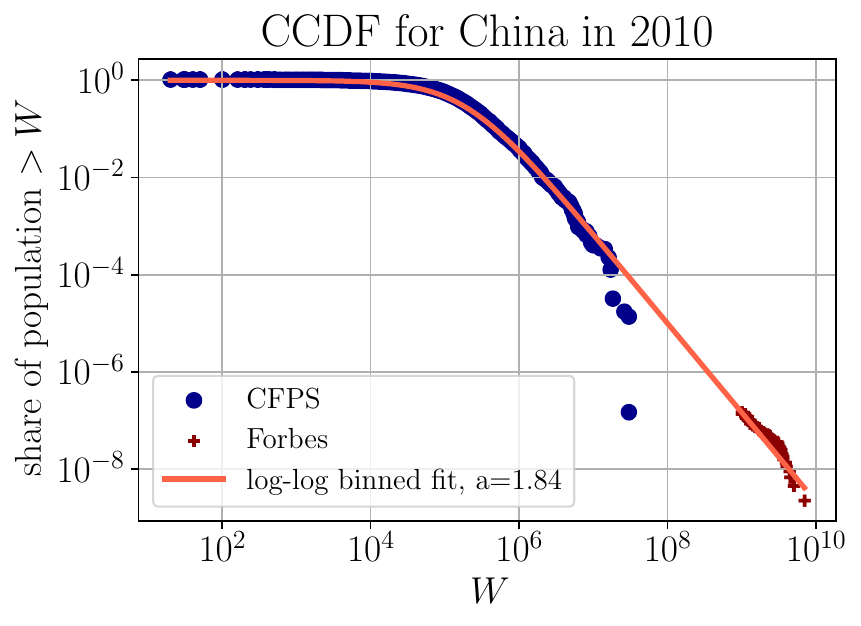}}%
 \quad
  {\includegraphics[width=2.2in]{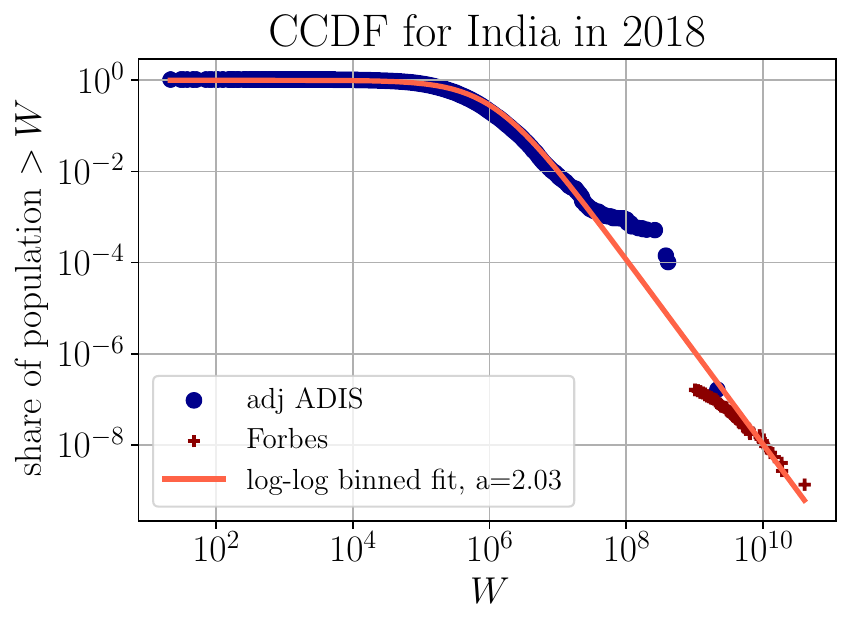}}%
 \quad
 {\includegraphics[width=2.2in]{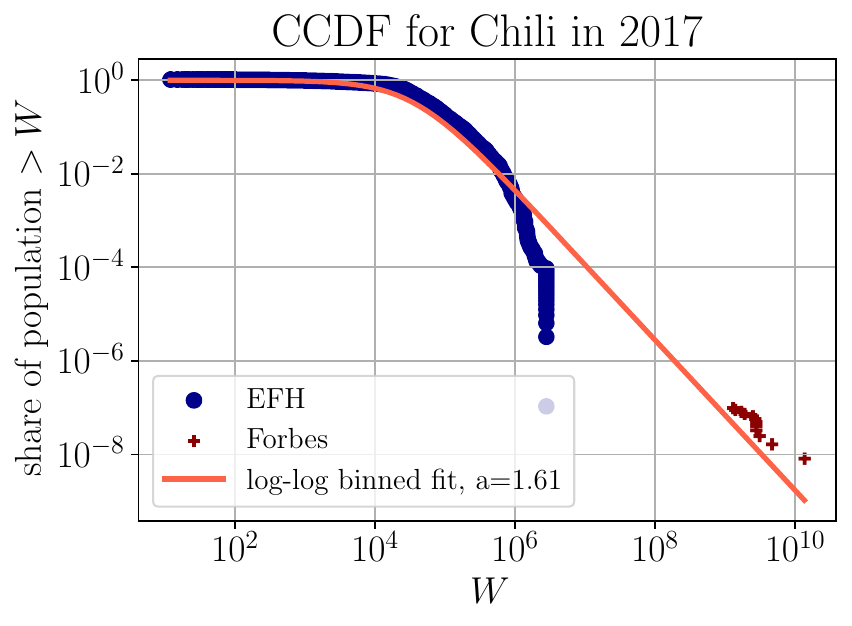}}%
 \quad

 \caption{CCDF for different countries in the world, aggregating rough data from national surveys and Forbes'billionaires list. The $a$-value corresponds to the exponent of the fitted power-law.
 }%
 \label{fig:wealth_empirical}
\end{figure*}

\begin{table*}[h!]
\caption{\label{tab:inequalities}Comparison of values of Gini and 1\% share of wealth we obtained.}
\begin{tabular}{lllll|llll}
                                                   & \multicolumn{4}{c|}{\textbf{Gini coefficient}}                                               & \multicolumn{4}{c}{\textbf{Top 1\% share}}                                                   \\  
                                                   & \textbf{Model} & \textbf{Emp(S+F)} & \textbf{Emp(S)} & \textbf{Official} & \textbf{Model} & \textbf{Emp(S+F)} & \textbf{Emp(S)} & \textbf{Official} \\ \cline{2-9} \noalign{\vskip 0.85mm} \cline{2-9}
                                                   
\multicolumn{1}{l||}{\textbf{United States (2022)}} & 84.2                 & 85.3                     & 82.8                   & 83.0              & 51.8                 & 44.1                     & 34.8                   & 35.1              \\
\multicolumn{1}{l||}{\textbf{France (2021)}}        & 78.4                 & 69.4                     & 67.6                   & 70.2              & 38.4                 & 21.6                     & 16.8                   & 22.3              \\
\multicolumn{1}{l||}{\textbf{Spain (2021)}}         & 74.5                 & 68.7                     & 67.9                   & 69.1              & 30.9                 & 24.1                     & 22.1                   & 23.1              \\
\multicolumn{1}{l||}{\textbf{Italy (2021)}}         & 74.5                 & 67.8                     & 67.0                   & 67.2              & 30.9                 & 22.0                     & 20.2                   & 23.3              \\
\multicolumn{1}{l||}{\textbf{Germany (2021)}}       & 79.4                 & 74.0                     & 72.7                   & 78.8              & 40.6                 & 21.6                     & 17.8                   & 31.7              \\
\multicolumn{1}{l||}{\textbf{China (2010)}}         & 68.7                 & 66.1                     & 66.0                   & 70.0              & 21.6                 & 18.4                     & 18.3                   & 31.5             
\end{tabular}
\captionsetup{justification=raggedright} % Centrage du texte de la deuxième légende
\caption*{\textbf{NB:} The \textit{Model} values are computed from the estimated Pareto exponent of our method. This Pareto exponent, unlike the other values, is computed on net wealths $W \geq 1$ only. This creates a bias and could explain some differences with the other values. \textit{Emp} values are computed empirically on all datapoints: (S+F) for national Survey + Forbes, (S) for national Survey only. Last, the \textit{Official} values are extracted from the \textit{Global Wealth Databook 2023}, by UBS and Credit Suisse.}
\end{table*}
Following the methodology proposed by Vermeulen~\cite{vermeulen2018fat}, we conducted an empirical study of wealth distributions by aggregating data from Forbes' billionaires list with national wealth surveys. Specifically, we used the HFCS for European countries, the SCF for the United States, the CFHS for China, the AIDIS for India (adjusted following Malhotra~\cite{DVN/TMKGOU_2024}), and the EFH for Chile. Combining these heterogeneous sources is non-trivial, as a sizable fraction of the population—typically those with net wealth $W$ between $10^7$ and $10^9$—is captured neither by national surveys, which tend to under-sample the very wealthy, nor by the Forbes list, which only reports the top of the distribution. This sampling gap manifests itself as a slight discontinuity around $W \approx 10^8$ in most of the empirical CCDFs reported in Fig.~\ref{fig:wealth_empirical}. The United States constitutes a notable exception, where this artifact is barely visible, consistent with the lower nonresponse bias of wealthy households in the SCF documented by Bricker et al.~\cite{BRICKER2019108659}.

In what follows, we restrict the analysis to strictly positive net wealth ($W \geq 1$) and focus on the upper tail of the resulting distribution. To estimate the Pareto exponent in a robust way, we adopt a logarithmic-binning procedure in the fitting, so that the dense set of points originating from national surveys does not disproportionately weight the regression in log-log space. The resulting CCDFs, displayed in Fig.~\ref{fig:wealth_empirical} for several countries, allow us to characterize the upper tail of the wealth distribution and to extract a country-specific Pareto exponent. Using the closed-form expressions derived in Section~\ref{sec:init}, these exponents can in turn be mapped onto theoretical values of the Gini coefficient and of the wealth share held by the top percentiles. We compare these model-implied values with their empirical counterparts and with the estimates reported in the \textit{Global Wealth Report 2023} of Credit Suisse\cite{suisse2023global}; the results are summarized in Table~\ref{tab:inequalities}. 

\clearpage
\section{Pseudo-code}
\label{Pseudo_code}

\begin{figure*}[h!]
\label{Pseudo_code_policy}
\begin{minipage}{\linewidth}
\begin{algorithm}[H]

  \caption{ABM\_Compute with Policy 6}
    % \label{ABM_Compute_with_policy}
   \begin{algorithmic}
   % \algsetup{linenosize=12}
   %METTRE LISTE DES PARAMETRES UTILISES
   \Require $N=10k$, $\lambda=50\%$, $Gini_0$, $ratio_\mathrm{G} = 15\%$, $r_0 = 7\%$, $I = 5\%$, $r_{\mathrm{loss}}=10\%$, $k_{\tau} = \frac{2}{\tau+1}$, $k_{\theta} = \frac{2}{\theta+1}$, $W_\mathrm{max} = 100$, $a = 2.15$, $\varphi_\mathrm{im}=0.1\%$
   \Ensure $\mathbf{W_{\mathrm{G}}}$ and $\mathbf{W_{\mathrm{B}}}$
   \State $ \mathbf{W_\mathrm{G}}[0] \gets \mathbf{W}_{initial}\times ratio $ \Comment{$W_{\mathrm{G}}$ is a $(t_\mathrm{max} \times n)$ matrix}
   \State $\mathbf{W_\mathrm{B}}[0] \gets \mathbf{W}_{initial}\times (1-ratio)$ \Comment{$W_{\mathrm{B}}$ is a $(t_\mathrm{max} \times n)$ matrix}
   \State $Y[0] \gets (0)_{n}$ \Comment{Y is $(t_\mathrm{max} \times n)$ matrix, incomes of agents}
   \State $\mathrm{EMA\_WB}_{k3}[0] \gets k_{\tau} \times \sum_i W_\mathrm{B}\left[0\right]\left[i\right]$ \Comment{$(t_\mathrm{max})$ vector, for computation of returns}
   \State $\mathrm{EMA\_WB}_{k100}[0] \gets k_{\theta} \times \sum_i W_\mathrm{B}\left[0\right]\left[i\right]$ \Comment{$(t_\mathrm{max})$ vector, for computation of shock's probability}
   \State $\alpha_\mathrm{eff}[0] \gets (0)_n$ \Comment{$\alpha_\mathrm{eff}$ is a $(t_\mathrm{max} \times n)$ matrix, effective tax rate factor}
   \State $\alpha_\mathrm{credit}[0] \gets 1$ \Comment{Credit tax (refunded to tax payers)}
   \State $r_\mathrm{boost}[0] \gets 0$ \Comment{Green-boost rate (given to Green investors)}
   
   \For{time $t \in [0, t_\mathrm{max}]$}
        \State $r_\mathrm{G}[t] \gets r_0 + \left(\frac{1}{2} - \frac{\mathrm{EMA\_WB}_{k3}[t]}{\sum_i W[t][i]}\right)$
        \State $r_\mathrm{B}[t] \gets r_0 + \left(\frac{1}{2} + \frac{\mathrm{EMA\_WB}_{k3}[t]}{\sum_i W[t][i]}\right)$
        \State $P_{\mathrm{BOOM}} \gets \frac{1}{2}\left(1 + \mathrm{tanh}\left(\frac{\mathrm{EMA\_WB}_{k100}[t]}{a} -b \right)\right)$ 
        \State $\delta y_\mathrm{tax} \gets 0$ \Comment{temp. variable to store the tax paid by Brown investors}
        \State $y_\mathrm{G} \gets 0$ \Comment{temp. variable to store income of Green investors}
        \For{each agent $i \in [1,N]$}
            \State $\rho[t][i] \gets \mathrm{rank}(\mathbf{W}[t][i]|\mathbf{W}[t])$ \Comment{Rank of agent $i$'s wealth}
            \State $Y[t][i] \gets r_\mathrm{G}[t]W_\mathrm{G}[t][i]+r_\mathrm{B}[t]W_\mathrm{B}[t][i]$
            \State $C_\mathrm{B}[t][i] \gets r_{\mathrm{loss}} \times \frac{1}{2}\left(1 + \mathrm{tanh}\left(\frac{\mathrm{EMA\_WB}_{k100}[t]+k_{\theta}\times Y[t][i]}{a} - b \right)\right)$
            \State $C_\mathrm{G}[t][i] \gets r_{\mathrm{loss}} \times \frac{1}{2}\left(1 + \mathrm{tanh}\left(\frac{\mathrm{EMA\_WB}_{k100}[t]}{a} - b \right)\right)$
            \State $\Delta U[t][i] \gets \frac{1-\lambda}{\lambda_a}(r_\mathrm{G}[t] + r_\mathrm{boost} - r_\mathrm{B}[t] + r_\mathrm{tax}\alpha_\mathrm{credit}[t]\alpha_\mathrm{eff}[t][i]) \times \frac{Y[t][i]}{\sum_i Y[t][i]} - \frac{\lambda}{\lambda_b}\mathrm{F}(\rho[t][i]) \left(C_{\mathrm{G}}[t][i] - C_{\mathrm{B}}[t][i]\right)$
            
            \State $\alpha_\mathrm{eff}[t][i] \gets \mathrm{compute\_effective\_taxrate}(Y[t][i] | Y[t])$ \Comment{See Fig.~\ref{fig:effective_rate} and Eq.~\eqref{eq:alpha_eff}}
            \If{$\Delta U[t][i] < 0$} agent $i$ chooses B:
                \State $\delta y_\mathrm{tax} \gets \delta y_\mathrm{tax} + Y[t][i]\times \alpha_\mathrm{eff}[t][i]\times r_\mathrm{tax}$
            \Else \ agent $i$ chooses G:
                \State $y_\mathrm{G} \gets y_\mathrm{G} + Y[t][i]$
            \EndIf
        \EndFor
    
        \State $r_\mathrm{boost}[t+1] = \frac{\delta y_\mathrm{tax}}{\sum_iY[t][i]}$
        \State $\alpha_\mathrm{credit}[t+1] = \frac{y_\mathrm{G}}{\sum_iY[t][i]}$

        \State rand $\gets Uniform(0,1)$
        \If{rand $< P_{\mathrm{BOOM}}[t]$}
            \For{each agent $i \in [1,N]$}
                \State $\mathrm{rand}_\mathrm{loss} \gets Uniform(0,1)$
                \State $W_\mathrm{B}[t+1][i] \gets W_\mathrm{B}[t+1][i]\times (1 - 2r_{\mathrm{loss}}\cdot \mathrm{rand}_\mathrm{loss})$
                \State $W_\mathrm{G}[t+1][i] \gets W_\mathrm{G}[t+1][i]\times (1 - 2r_{\mathrm{loss}}\cdot \mathrm{rand}_\mathrm{loss})$
            \EndFor
            
        \EndIf
        
        \For{each agent $i \in [1,N]$}
            \If{$\Delta U[t][i] < 0$} agent $i$ chooses B:
                \State $W_{\mathrm{B}}[t+1][i] \gets W_{\mathrm{B}}[t][i](1+r_{\mathrm{B}}[t](1-\alpha_\mathrm{eff}[t][i]\alpha_\mathrm{credit}[t+1]r_\mathrm{tax})-a_{\mathrm{B}}) + W_{\mathrm{G}}[t][i]r_{\mathrm{G}}[t](1-\alpha_\mathrm{eff}[t][i]\alpha_\mathrm{credit}[t+1]r_\mathrm{tax})$
                \State $W_{\mathrm{G}}[t+1][i] \gets W_{\mathrm{G}}[t][i](1-a_{\mathrm{G}})$
            \Else \ agent $i$ chooses G:
                \State $W_{\mathrm{B}}[t+1][i] \gets W_{\mathrm{B}}[t][i](1-a_{\mathrm{B}})$
                \State $W_{\mathrm{G}}[t+1][i] \gets W_{\mathrm{G}}[t][i](1+r_{\mathrm{G}}[t]\times(1+r_\mathrm{boost}[t+1])-a_{\mathrm{G}}) + W_{\mathrm{B}}[t][i]r_{\mathrm{B}}[t]\times (1+r_\mathrm{boost}[t+1])$
            \EndIf
        \EndFor

        \State $\mathrm{EMA\_WB}_{k100}[t+1] = k_{\theta} \sum_i W_{\mathrm{B}}[t+1][i] + (1-k_{\theta})\mathrm{EMA\_WB}_{k100}[t]$
        \State $\mathrm{EMA\_WB}_{k3}[t+1] = k_3 \sum_i W_{\mathrm{B}}[t+1][i] + (1-k_3)\mathrm{EMA\_WB}_{k3}[t]$
    \EndFor

   \end{algorithmic}
\end{algorithm}
\end{minipage}
\end{figure*}

\end{document}